\documentclass[aps,prl,superscriptaddress,floatfix,amsmath,amsfonts,twocolumn]{revtex4-2}  

\usepackage{amsfonts}
\usepackage{upgreek}
\usepackage{braket}
\usepackage{verbatim}
\usepackage{xr}
\usepackage{enumitem}
\usepackage{amsmath}
\usepackage{graphicx}
\usepackage{epsfig}
\usepackage{bm}
\usepackage{bbold}
\usepackage{amssymb}
\usepackage{color}
\usepackage[normalem]{ulem}  %

\newcommand{\micro}{${\upmu}$}

\newcommand{\um}{$\,$\micro m}
\newcommand{\uev}{$\,$\micro eV}

\begin{document}

\title{Phase locking of ring-shaped exciton-polariton condensates to coherent optical drive } 

\author{A.N. Osipov}
\affiliation{School of Physics and Engineering, ITMO University, Saint Petersburg 197101, Russia}
\affiliation{Center for Theoretical Physics of Complex Systems, Institute for Basic Science (IBS), Daejeon 34126, Republic of Korea}

\author{S.V. Koniakhin} \thanks{kon@ibs.re.kr}
\affiliation{Center for Theoretical Physics of Complex Systems, Institute for Basic Science (IBS), Daejeon 34126, Republic of Korea}
\affiliation{Basic Science Program, Korea University of Science and Technology (UST), Daejeon 34113, Republic of Korea}
\author{O.I. Utesov}
\affiliation{Center for Theoretical Physics of Complex Systems, Institute for Basic Science (IBS), Daejeon 34126, Republic of Korea}

\author{I.S. Aronson}
\affiliation{Departments of Biomedical Engineering, Chemistry, and Mathematics,
The Pennsylvania State University, University Park 16802, USA}

\author{A.V. Yulin}
\affiliation{School of Physics and Engineering, ITMO University, Saint Petersburg 197101, Russia}

\date{\today}

\begin{abstract}

The effect of an additional quasi-resonant drive on the dynamics of the ring-shaped incoherently pumped polariton condensates carrying angular momentum (vorticity) is studied theoretically. Numerical simulations of the 2D and 1D Gross-Pitaevskii equations show that the difference of the topological charges(vorticities) $\Delta n$ of the condensate and the quasi-resonant coherent drive plays a crucial role in the synchronization dynamics. It is shown that in an axially symmetric system, synchronization can only occur if $|\Delta n| = 0$, whereas in the other cases the phase of the condensate cannot be locked to the phase of the coherent drive. To explain this effect observed in the numerical simulations a perturbation theory is developed. The theory shows that the phase slip between the condensate and the coherent drive can be understood in terms of the motion of 2$\pi$ kinks. It is shown that the breaking of the axial symmetry can stop the motion of the kinks, allowing the phase locking of the condensate to the coherent drive.

\end{abstract}

\maketitle

\section{Introduction}
Exciton-polaritons are hybrid light-matter excitations emerging in microcavities with embedded quantum wells in the regime of strong light-matter coupling \cite{sanvitto2012exciton}. The interest to these excitations is driven by their rich dispersive and nonlinear properties allowing to observe such interesting phenomena as Bose-Einstein condensation at unusually high temperature \cite{kasprzak2006bose}, optically and electrically driven polariton lasers \cite{butov2007polariton, schneider2013electrically}, Berezinskii-Kosterlitz-Thouless phase transition \cite{nitsche2014algebraic}, Zitterbewegung effect \cite{lovett2023observation}. It should also be noted that the polariton system look promising from practical point of view and a number of possible application have already been suggested  \cite{liew2011polaritonic, sanvitto2016road,berloff2017realizing, lagoudakis2017polariton, zasedatelev2019room, harrison2022solving, barrat2023superfluid}.

One of the challenging problems in polaritonics is the engineering of the states of the polariton condensates, in particular to control their frequencies and phases. This can be achieved by phase locking of the condensates to external coherent drive \cite{kalinin2018simulating, chestnov2019optical} or by mutual synchronization of the polaritons leading to the coherent state of many polariton condensate droplets \cite{ohadi2016nontrivial, berloff2017realizing, harrison2022solving, lagoudakis2017polariton}. The inter-condensate interaction through the evanescent fields and through the propagating polaritons have been studied theoretically and experimentally \cite{abbarchi2013macroscopic, ohadi2016nontrivial, topfer2020time}. In Ref. \cite{ohadi2016nontrivial} it is shown that in the latter case the mutual phase of the interacting identical condensates can be either $0$ (ferromagnetic state) or $\pi$ (anti-ferromagnetic state). In the case of long inter-condensate distances, the coupling cannot be considered as instantaneous and the delay time affects the condensates dynamics \cite{topfer2020time}.

The interplay between mutual synchronization of the condensate droplets and the synchronization to external drive has been investigated theoretically \cite{kalinin2018simulating, chestnov2019optical}. It was demonstrated that this complex interaction drastically affects the polariton dynamics, in particular, in Ref.~\cite{chestnov2019optical} it is shown that spontaneous symmetry breaking can occur resulting in the formation of a stable state with the mutual phase of the condensates not equal neither to $0$ nor to $\pi$.



Another problem, attracting significant attention of scientific community during the last decade, is the engineering of the angular momentum of light, including vortices in exciton-polariton systems. They are topologically protected states characterized by nontrivial phase winding around singular core (topological charge) \cite{pismen1999vortices}.  Exciton-polariton vortices have been widely studied experimentally and theoretically \cite{dall2014creation, dreismann2014coupled, sigurdsson2014information, ma2016incoherent, ma2020realization, gnusov2023quantum, PhysRevB.105.L060502}. The arrays of interacting vortices have been studied theoretically \cite{keeling2008spontaneous, chestnov2021dissipative, harrison2023polariton} and experimentally  \cite{tosi2012geometrically, gao2018controlled, alyatkin2024antiferromagnetic} including the vortices forming in periodical potentials where vortices-like geometrical frustrations were recently observed \cite{cookson2021geometric}.

The Mexican-hat complex potentials created by incoherent pumps open a possibility to create stable ring-shaped condensates with different topological charges. There are many ways to control their vorticity \cite{dall2014creation, ma2020realization, gnusov2023quantum} including the use of a coherent drive transferring its vorticity to the condensate \cite{sigurdsson2014information}. Moreover, there are some experimental evidence, that topological charge of non-coherent pumping could also affect polariton condensate's vorticity \cite{kwon2019direct}.




The aim of the present work is to investigate the effect of a resonant coherent drive on the polariton condensate formed via the condensation from a reservoir of incoherent excitons created by external pumping (so-called incoherent drive).  We restrict our consideration to the case of weak coherent drive that does not change the condensate density much but nevertheless can cause nontrivial dynamics of the condensate phase distribution. As it will be shown below, the synchronization of the spatially distributed condensates depends drastically on the topological charges (vorticity) of the drive and the condensate. 


The paper is organized as follows. In the next section we introduce a mathematical model describing polariton dynamics in mean-field approximation by two-dimensional Gross-Pitaevskii equation (GPE). We also consider the case of strong confinement of the condensate along the radial coordinate allowing to assume that the radial distribution of the condensate is defined mostly by the effective potential created by the pump. Then it is possible to reduce the mathematical model to a one-dimensional Gross-Pitaevskii equation describing the evolution of the condensate along the angular coordinate. 

The third section of the paper is devoted to numerical modeling of the 2D condensate dynamics. We examine the behavior of the condensate under the action of the coherent drive. We study the synchronization for the case, when condensate and coherent drive have different topological charges. We show that, in this case, the angular distribution of the condensate phase has a shape of a moving kink and that no synchronization occurs if the potential and the drive intensity are axially symmetric. Let us remark that the kink-like phase distributions are well known for Bloch waves forming in non-equilibrium nonlinear system described by complex Ginzburg-Landau equation \cite{coullet1992strong}, and has been studied later in the context of spontaneous symmetry breaking \cite{coullet1990breaking} and the formation  spiral waves \cite{aranson1995frequency}. Let us also remark that the discussed kink solution can also exhibit chaotic dynamics \cite{aranson1985development}.

In section IV of this paper we consider in detail the effect of kink motion on the dynamics of the polariton condensates for the case where the condensate has a shape of a ring-shaped belt with the width of the belt much smaller compared to its radius. Strong localization along the radial coordinate can be provided by an effective potential created either by sample microstructuring or by an appropriately shaped incoherent pump. This makes it possible to assume that the radial structure of the condensate is dictated by the potential and reduce the problem to a much simpler 1D one. Using this model we reproduce the effect discussed in the section III and study different regimes of the pinning of the kinks by the inhomogeneities of the effective potential (the shape of the potential is fixed along the radius but it can depend on the angle). A perturbation theory is developed for the case when the coherent drive is so weak that a linear equation can be used to describe the density perturbations caused by it. However, in this case, the dynamics of the condensate phase remains strongly nonlinear in this case provided that the size of the condensate is large enough. We derive the equation governing the temporal evolution of the condensate phase and show that in the limiting case this equation reduces to a well-known overdamped sine-Gordon equation.

This allows us to treat the dynamics of such ring-shaped condensates in terms of moving kinks having the shapes of resting solution of the sine-Gordon equation. Following this strategy we develop a quasi-particle approach in the section V and derive ordinary differential equation for the positions of the kinks. This equation can be solved analytically even for some cases where the axial symmetry is broken by an effective potential. We demonstrate that this quasi-particle approach gives an accurate description of different regimes of synchronization.



Finally, in conclusion we briefly summarize the main results of the paper and give some outlook for further research.

\section{Model}
The dynamics of the exciton-polariton condensate within the mean field approximation could be efficiently described by 2D driven-dissipative Gross-Pitaevskii equation \cite{wouters2007excitations}: 

\begin{subequations}
    \begin{eqnarray}
        &&i\hbar \partial_\tau\Psi = \bigg[-\frac{\hbar^2\hat{k}^2}{2m_p}(1-i\Lambda)+ g_c|\Psi|^2 + g_R n_R(\bm{r}, t)  + \nonumber\\
        && + i\frac{\hbar}{2}(R n_R(\bm{r}, t) - \gamma_c)\bigg]\Psi + P_c(\bm{r}, t), \\
        &&\partial_\tau n_R = -(\gamma_R + R|\Psi|^2)n_R(\bm{r}, t) + P_R(\bm{r}),
        \label{eq_GP}
    \end{eqnarray}
\end{subequations}
where $m_p$ - the polariton effective mass, $g_c$ - polariton-polariton interaction strength, $g_R$ - shift of the condensate frequency due to the interaction with reservoir, $\gamma_c, \gamma_R$ - polariton and reservoir relaxation rates, R - rate of the condensation of the reservoir particles, $\Lambda$ is the energy relaxation parameter \cite{solnyshkov2014hybrid, cookson2021geometric}, $P_R$ - incoherent (nonresonant) pump of the reservoir, $P_c = f_0e^{il_f\theta - i\delta t }\Theta(t-t_0)$ - coherent (resonant) pump where $t_0$ - time when coherent pump is switch on, $l_f$ - coherent pump's vorticity.

To obtain ring-shaped condensation with a big radius two pump forms could be used. First, condensation in the pump region under the intensive single ring pump $P_R(r) = P_0 e^{-\frac{(r-r_0)^2}{w^2}}$. This case is earthier to realize experimentally and numerically \cite{cookson2021geometric, ma2018vortex, ma2020realization} which we have done in the next section of the paper. However, the 2D simulations of the Gross-Pitaevskii equation are time consuming and the complexity of the system make it difficult to distill the essence of the physical process responsible for the dynamics observed in the simulations.

Therefore, to shed more light on the discovered phenomena we introduce a simpler but physically relevant effective 1D model assuming that the motion of a condensate along one direction is suppressed by a deep external potential. This can be done, for example, by the microstructuring of the sample creating the deep traps for the polaritons. Let us note that a similar potential can also be produced by spatially non-uniform incoherent pump.  

To proceed, we eliminate equation for the reservoir which is possible under the assumption that the reservoir dynamics is fast compared to the characteristic time scales of the polariton condensate dynamics. Assuming that the reservoir density follows the polariton density adiabatically, we obtain $n_R =\frac{ P_R(\bm{r})}{\gamma_R + R|\Psi|^2}$. For the condensates of relatively low densities $|\Psi|^2 \ll \frac{\gamma_R}{R}$ the expression for the $n_R$ can be expanded in Taylor series and, keeping the two leading terms, we obtain the Gross-Pitaevskii equation for the condensate order parameter in the form    
\begin{eqnarray}
        &&i\hbar \partial_\tau\Psi = \Bigg[-\frac{\hbar^2\hat{k}^2}{2m_p}(1-i\Lambda)+ \left(g_c-\frac{P_R R}{\Gamma_R^2}\right)|\Psi|^2 +\tilde V  + \nonumber\\
        && + i\frac{\hbar}{2}\left(R\frac{P_R}{\Gamma_R}\left(1 - \frac{R}{\gamma_R}|\Psi|^2\right)- \gamma_c\right)\Bigg]\Psi + P_c(\bm{r}, t). 
        \label{eq_GP_adiabatic}
\end{eqnarray}
where $\tilde{V} = g_R P_R/\Gamma_R$ is the effective potential created by the incoherent pump.

For our purposes we assume that the potential is a ring-shaped axially symmetric well. The exact form of the potential does not crucial for our considerations, however for the sake of concreteness we consider its given by the pump $P_R(r) = P_0 (e^{-\frac{(r-r_0 + d/2)^2}{w^2}} + e^{-\frac{(r-r_0 - d/2)^2}{w^2}})$. For this type of potential, condensation occurs in the trap created by two pump peaks, not in the pump region as in the first case. We assume that the potential $\tilde V$ is that deep that it provides tight confinement of the polaritons along the radius so that only the lowest radial mode is of importance. Then an effective 1D Gross-Pitaevskii equation can be derived to describe the evolution of the condensate in such a potential \cite{chestnov2021dissipative, munoz2020long}. Using the latter model, we study the evolution of the condensate under the action of the external coherent drive in detail. We also develop a perturbation theory capturing the principal features of the dynamics.  


In the case of strong radial confinement in the trap, it is possible to separate radial and the angular dependencies representing the order parameter as $\Psi = F(r)\psi(\theta)$ \cite{cherotchenko2021optically}.  Then multiplying the equation (\ref{eq_GP_adiabatic}) by $F^*(r)$ and integrate over the radial coordinate, we obtain an effective 1D Gross-Pitaevskii equation for the angular part of the wavefunction
\begin{equation}
    \partial_{\tau} \psi = \left((\gamma-\Gamma|\psi|^2) + i \delta - iU|\psi|^2 +iJ\partial_{\theta\theta} \right)\psi - i fe^{il_f\theta},
    \label{1D GP dim}
\end{equation}
where $\gamma$ is a trapped mode growth rate given by the solution of the non-hermitian eigenproblem \cite{cherotchenko2021optically}, $\delta$ is a detuning of resonant pumping from the trapped mode frequency, $\Gamma$ is a nonlinear gain saturation parameter,  $U$ is an effective polariton-polariton interaction coefficient, $J$ is an effective angular dispersion along the ring. The last term in the equation corresponds to the resonant pumping with amplitude $f$ and topological charge $l_f$.
All the parameters can be found by calculations of the corresponding integrals (see Appendix \textbf{A}) or, alternatively, they can be found by fitting the results of the simulations of 1D model to the experiment or to the simulations of the initial 2D model.

For the sake of mathematical convenience, it is worth writing Eq. (\ref{1D GP dim}) in a dimensionless form
\begin{equation}
    \partial_{t} \psi' = \left(1-|\psi'|^2 + i \delta' - i\alpha|\psi'|^2 +iJ'\partial_{\theta\theta} \right)\psi' - i f'e^{il_f\theta},
    \label{1D GP}
\end{equation}
where $t = \tau/\gamma$, $J' = J/\gamma$, $\alpha = U/\Gamma$ is the ratio between polariton - polariton interaction strength and nonlinear gain saturation \cite{PhysRevB.85.235303}, $\delta' = \delta/\gamma$, $\psi' = \psi\sqrt{\Gamma/\gamma}$, $f' = f\sqrt{\Gamma/\gamma}$. In what follows, to make the notation more brief, we will omit the dashes. Note, that derived 1D Gross-Pitaevskii equation is also known as 1D Complex Gindzburg-Landau equation which is known to exhibit rich dynamics including chaotic and time periodic solutions \cite{schopf1991small}.

Let us remark that the applicability of the derived 1D equation is limited by the conditions imposed above. However the qualitative agreement between the results obtained by the full scale modeling of Eq. (\ref{eq_GP}) and the simulations of Eq. (\ref{1D GP}) takes place even beyond the applicability range of (\ref{1D GP}). Thus the simpler model gives a clues allowing to understand a more complex dynamics of 2D polariton condensates and, in particular,  to provide a good qualitative understanding of synchronization effects discussed below.

Let us also remark that the same 1D equation can be derived as a limiting case (long-wave limit) of the coupled Stuart-Landau equations \cite{lagoudakis2017polariton} describing a periodic chain of the condensates droplets \cite{cookson2021geometric}
\begin{eqnarray}
    &&\partial_t \psi_i = \left((1-|\psi_i|^2) + i \delta - i\alpha|\psi_i|^2 + i2J\right)\psi_i + \nonumber \\
    &&+i J(\psi_{i+1} + \psi_{i-1} -2 \psi_i) 
    - i fe^{il_f\theta_i}
    \label{2D GP discrete}.
\end{eqnarray}

\section{Numerical simulations of 2D ring condensate dynamics in the presence of coherent pumping}

We start with the modeling of 2D problem (\ref{eq_GP}) with the pump in a form of a ring with the gaussian radial profile $P_R(r) = P_0 e^{-\frac{(r-r_0)^2}{w^2}}$ which provides the condensation of the polaritons from exciton reservoir. We use the most general model containing the equations for the polariton and the reservoir. We assume that the sample is spatially uniform and thus the conservative potential is absent. The advantage of such setups is that they can relatively easily be realized in experiments \cite{cookson2021geometric, ma2018vortex, ma2020realization}.  

Numerical simulations reveal that in a wide range of parameters multistability takes place. If the initial conditions are taken in a form of weak noise then the stationary state can have different winding numbers what matches the results reported earlier in a number of papers \cite{cookson2021geometric, sigurdsson2014information, ma2020realization, gnusov2023quantum}. The probability of the formation of a state with a given topological charge depends on the absolute value of the charge and on the size of the ring and on the intensity of the pump \cite{cherotchenko2021optically}. We consider the rings of relatively large sizes pumped well above the condensation threshold. It allows us to observe the formation of the states with different angular indexes $l_c = 0, 1, 2, 3...$. 

\begin{figure}[!t]
\begin{center}
\includegraphics[width=\linewidth]{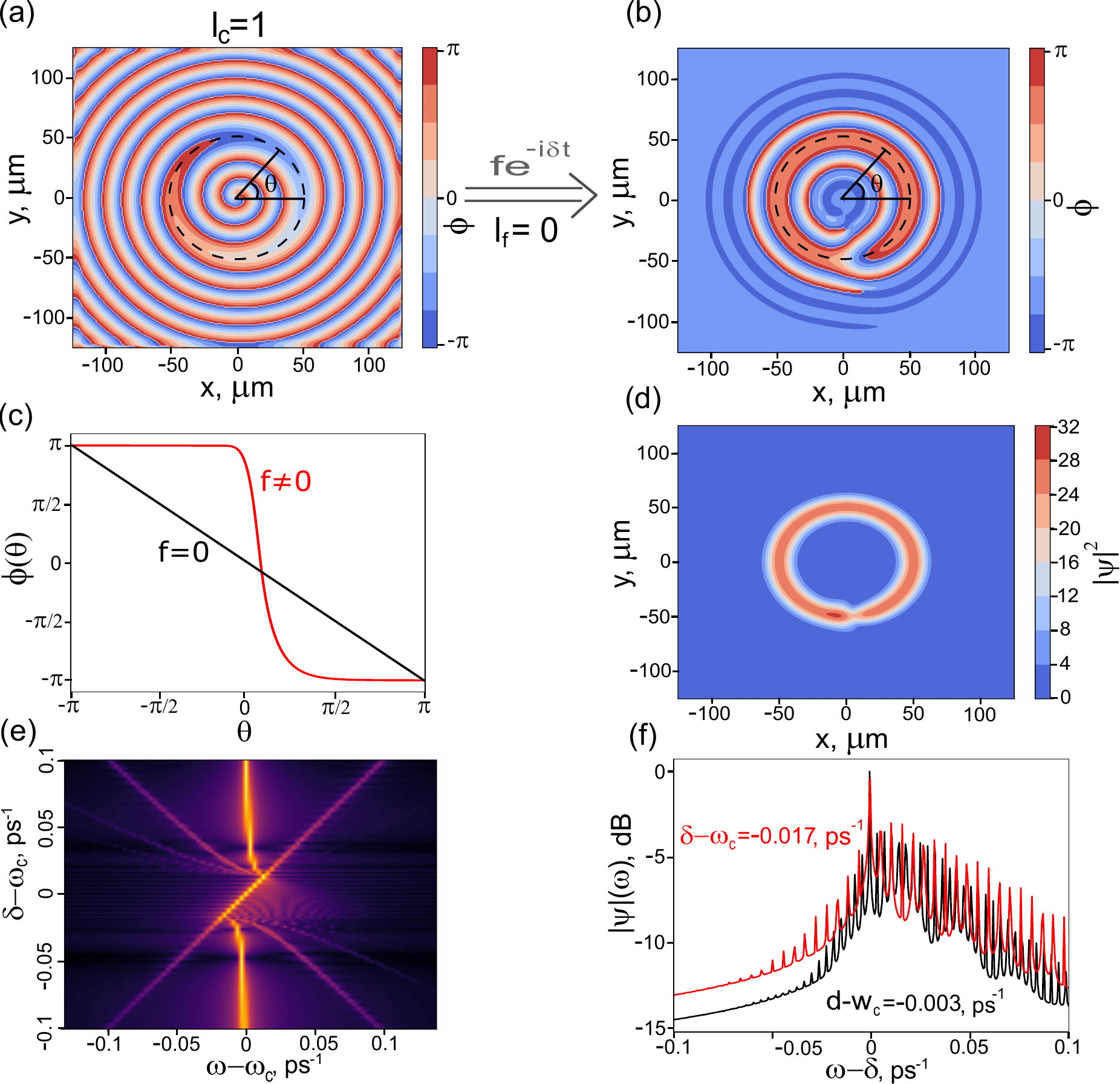}
\end{center}
\caption{ \label{fig1}
(a) The phase of the ring-shaped condensate with angular momentum $l_c = 1$ before applying the coherent pumping (b) The phase of the ring-shaped condensate with angular momentum $l_c = 1$ after turning on the coherent pumping with angular momentum $l_f = 0$ (c) Black and red solid lines are linear plots of the phase of the condensates (a) and (b) along the ring. (d) Condensate density corresponding to the phase shown in (b). (e) Spectum of exciton-polaritons on the ring for different coherent pump detunings $\delta$ with respect to condensate's frequency $w_c \approx 0.166$ ps$^{-1}$($\approx 0.11$ meV). (f) Spectrum for condensates under coherent pump with detunings $\delta-w_c = -0.003$ ps${^{-1}}$(black line) and $\delta-w_c = -0.017$ ps${^{-1}}$(red line). The spectra are shifted on coherent drive frequencies. The formation of frequency combs appears from the kink's movement evidencing the absence of synchronization. The frequency combs repetition rate and relative intensity with respect to pump line are determined by kink's speed. Hence, first detuning $\delta-w_c = -0.003$ ps${^{-1}}$ corresponds to slower kink compare to the other detuning, which was also observed directly from numerical simulations. The following parameters were used: $m_p = 0.35$ meVps$^2$\um$^{-2}$, $\gamma_c = 0.2$ ps$^{-1}$, $\gamma_r = 0.5$ ps$^{-1}$, $g_c = g_r = 3.3$ \uev\um${}^2$, $\hbar R = 33$\uev\um${}^2$, $\Lambda=0.1$, $w = 7.5$\um, $r_0 = 50$\um, $P_0 = 5P_{th}$, where $P_{th} = \frac{\gamma_R\gamma_c}{R}$; $\frac{f_0}{\hbar} = 0.048$ \um$^{-1}$ ps$^{-1}$.}
\end{figure}

The example of the condensate phase with winding $l_c = 1$, which appears to be the most probable state to grow from the initial noise, is illustrated in Fig.~\ref{fig1}(a), the modeling is done for experimentally achievable parameters given in the capture of Fig.~\ref{fig1}.  Let us also point out that to achieve a better control for the topological charge of the condensate an additional coherent beam can be used. Moreover, there are some indication that even the winding number of the incoherent pump can select the winding number of the condensate \cite{kwon2019direct}, though this effect is not captured by our model.   

After the condensate has formed, we applied a resonant pumping $P_c = f_0e^{il_f\theta - i\delta t }\Theta(t-t_0)$ with phase winding $l_f$. In the case where the resonant pumping and condensate have different vorticities $l_f \neq l_c$ and resonant pump frequency is close enough to the condensate's frequency (low detuning $\delta$), we observed that the difference of the condensate phase and the phase of the pump is not a linear function of the angle as it is in the case of the coherent pump of extremely low intensity, but form a step-like dependence, see Fig.~\ref{fig1}(a)-(c). The width of the step depends on the intensity of the coherent drive and thus in the trap of a given radius the pump must be sufficiently strong to produce a visible step of the phase. The dependence of the phase on the angle causes perturbation of the polariton density; the density perturbation is strongest at the area of the step (see Fig.~\ref{fig1}(d)). 

It is observed that for an axially symmetric incoherent pump the phase step is moving at a certain velocity. The velocity depends on the detuning $\delta$ between the free-running frequency of the condensate and the frequency of the coherent pump. This dynamics is observed for relatively small detunings, at larger detunings a complex phase dynamics has been observed (see Fig.~\ref{fig1} (e)).

Our simulations show that in annular geometry the condensate can be phase locked to a coherent drive only if the condensate and the coherent drive have equal topological charges. The absence of the synchronization can be explained in terms of the motion of the step of the mutual phase. If we have a look at the mutual phase dynamics at a certain point we will see that most of the time the mutual phase is constant but when the phase step passes the point the mutual phase slips by $2\pi$. Thus the mutual phase between the condensate and the coherent drive growth with the average rate $<\Delta \omega> = \frac{2\pi}{T}=\Omega $ where $T$ is the time that it takes for the phase step to make a full round trip, $\Omega$ is the angular velocity of the phase step.

The condensate spectrum has a form of frequency combs, see Fig.~\ref{fig1} (e), (f). Let us note that the most intense spectral line corresponds to the coherent drive frequency being not equal to the average frequency introduced above. We have checked that the spacing between the spectral lines in the comb is proportional to the angular velocity of the kink. This is why the density of the frequency comb depends on the detuning $\delta$ (see Fig.~\ref{fig1} (f)). Note that in the case when topological charges of the coherent drive and the condensate are equal, in our simulations phase locking of the condensate to the coherent drive takes place. The region of detunings where it happens seems to be similar to the region where we observed phase slips and frequency combs for the presented case.

In numerical simulations, we also observed that by breaking the rotational symmetry of the system, it is possible to pin the phase step on the inhomogeneities and achieve the phase locking between the condensate and the coherent pump. This configuration is studied in details in following sections. This naturally happens in discrete systems (\ref{2D GP discrete}) where the discreetness provides effective pinning provided  that the width of the phase step is comparable to the inter-sites distance. 

\section{1D dynamics of the ring condensate} 

To explain the nature of the observed effect, we consider a simpler system reducing the 2D problem to a one-dimensional annular system. For the sake of simplicity we consider the case where the reservoir dynamics can be adiabatically excluded allowing to use a 1D Gross-Pitaevskii equation (\ref{1D GP}) with periodic boundary conditions. 
For numerical simulations we take the initial conditions in the form of stationary solution of the GPE with incoherent pump only. This solution is $\psi(t=0) = \psi_0 e^{il_c\theta}$ and characterized by its amplitude $\psi_0=1$ (in our dimensionless units) and vorticity $l_c$. Then we study the evolution of this field under the action of the coherent drive. For mathematical convenience we introduce a new field $\tilde{\psi} = \psi e^{-il_f\theta}$ and write the equation in a reference frame rotating at the angular velocity $-J l_f$. This allows to remove the explicit dependencies of the coefficients on coordinates in equation \eqref{1D GP}. Let us remark that this change of variables preserve the structure of the equation but redefine the frequency detuning $\delta' = \delta - Jl_f^2$ and coherent pump vorticity $\tilde{l}_f = 0$. The initial condition in new variables takes the form $\psi(t=0) = \psi_0 e^{i\Delta_l\theta}$ where $\Delta_l = l_c-l_f$ is the difference of the topological charges of the pump and the initial condition. Evidently, the parameter defining the behavior of the system is the difference of the topological charges $\Delta_l$ but not their absolute values.

Now we proceed with the perturbation theory describing the dynamics of the condensate in terms of its phase. This approach, known as Kuramoto approximation \cite{kuramoto1984chemical}, works under the condition that the perturbation of the condensate density is small compared to its unperturbed value. In the presence of weak perturbations the density at each point is mostly determined by the balance between the pump and the losses with some relaxation rate. The perturbations can be considered to be small from the point of view of the density evolution if they are small compared to this relaxation rate. At the same time, in the leading approximation order, the dynamics of the phase depends on its spatial derivatives, regardless of how small these terms are. Therefore, the dynamics of the phase is governed by a partial differential equation which is, however, much simpler than the initial one because we can find the algebraic expression for the correction to the polariton density and  substitute it to the equation for the phase.   

It is convenient to write equation (\ref{1D GP}) in terms of condensate amplitude and phase $\psi = \rho e^{i\phi}$
\begin{subequations}
\begin{eqnarray}
    &&\partial_t \rho = (1-\rho^2)\rho - 2J\partial_\theta \rho\partial_\theta \phi - J \rho\partial_{\theta\theta} \phi - f\sin(\phi), \\
    \label{full eq for rho}
    &&\partial_t \phi = (\delta - \alpha\rho^2) - J(\partial_\theta \phi)^2 + J \frac{\partial_{\theta\theta}\rho}{\rho} - \frac{f}{\rho}\cos(\phi).
    \label{full eq for phi}
\end{eqnarray}
\end{subequations}

To proceed further, we assume big ring radius leading to $J \ll 1$,  small amplitude of a resonant pumping $f \ll 1$ and small detuning from the condensate frequency with respect to the blue shift $\tilde{\delta} = \delta - \alpha \ll 1$. All assumptions are given within our dimensionless units. Under these assumptions, the condensate density variation is very small and could be found as a perturbation  $\rho = 1 + \varepsilon$, where $\varepsilon \ll 1$. In the first order in the small parameters, density variation derived from Eq.(\ref{full eq for rho}) takes form
 \begin{equation}
     -2\varepsilon =J\partial_{\theta\theta}\phi + f \sin(\phi),
 \end{equation}
which, when substituting to Eq.(\ref{full eq for phi}), allows us to derive effective equation for the condensate phase:

\begin{equation}
    \partial_t \phi = \tilde{\delta} +\underbrace{\alpha (J \partial_{\theta\theta}\phi + f\sin(\phi))}_{\text{sine-Gordon}}  - J(\partial_\theta \phi)^2 - f\cos(\phi),
    \label{effective phase equation}
\end{equation}
where the highlighted part is nothing else but a textbook sine-Gordon equation's right part. Thus, for strong blue-shifts  $ \alpha \gg 1$, the stationary phase distribution of the condensate should be a solution of time-independent overdamped sine-Gordon equation. The boundary conditions for the solution is $\phi(0)=\phi(2\pi)+2\pi \Delta_l$.  If the length of the ring is large enough providing $J \ll 1$ then the solution can be seen as a train of $\Delta_l$ sine-Gordon kink solutions.

\begin{figure}[tb!]
\begin{center}
\includegraphics[width=\linewidth]{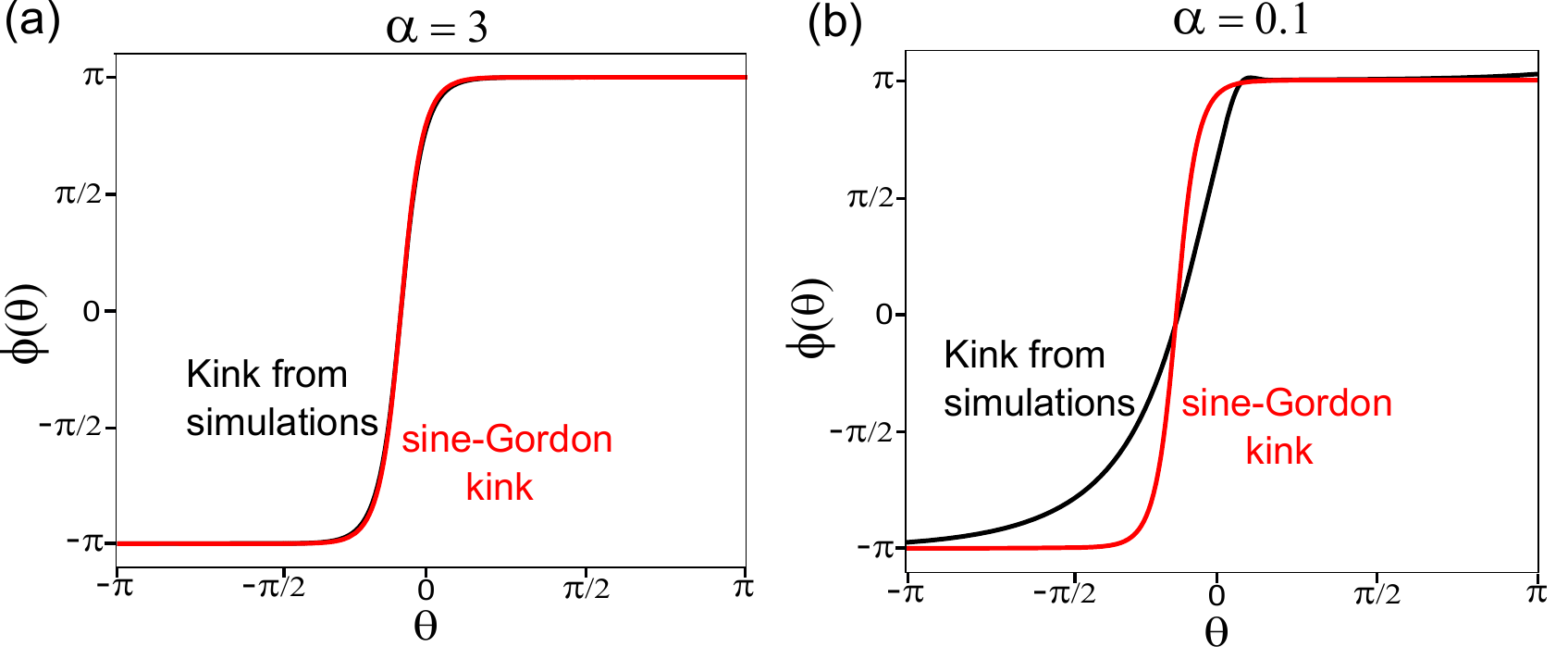}
\end{center}
\caption{ \label{fig2}
(a) and (b) Comparison of numerical phase profiles, obtained from simulations of 1D Gross-Pitaevskii equation (\ref{1D GP}), with sine-Gordon kinks given by equation $J \partial_{\theta\theta}\phi + f\sin(\phi) = 0$ for $\alpha = 3$ and $\alpha = 0.1$ respectively.}
\end{figure}

Let us note that the value $\alpha$ can vary depending on the microcavity. However, even for $\alpha = 3$ which is taken from \cite{PhysRevB.85.235303}, kink profiles are already similar to the sine-Gordon stationary kinks (see Fig.~\ref{fig2}(a)). Furthermore, in the case of small values of parameter $\alpha$ or even for $\alpha = 0$ the solution also have form of a kink changing only the shape, see Fig.~\ref{fig2}(b). For most of our simulations we took 
parameters $J = 0.00075, \, f = 0.04$, and $\alpha = 3$. In the following, these parameter values are assumed unless other parameters are explicitly stated.

\begin{figure}[tb!]
\begin{center}
\includegraphics[width=\linewidth]{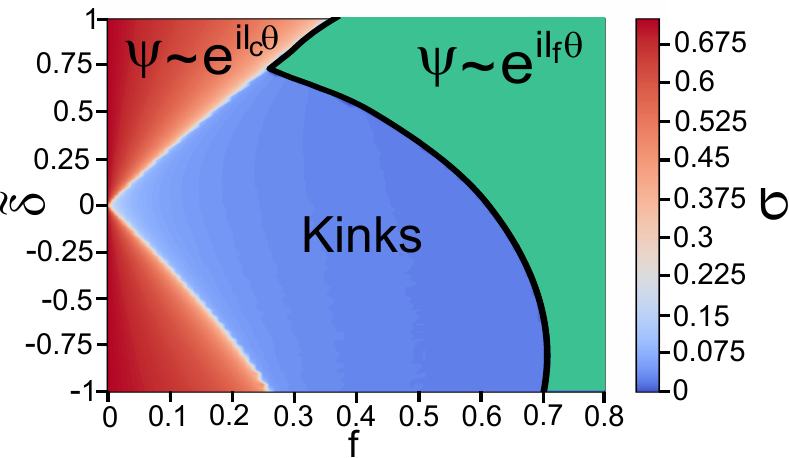}
\end{center}
\caption{ \label{fig3}
Phase diagram for the synchronization of ring condensate with resonant pumping. The quantity $\sigma = \sqrt{\int \partial_{\theta} \phi \theta^2 d\theta }/2\pi $ - ratio of the kink size calculated as a second momentum of phase $\theta$ derivative to the ring size($2\pi$) indicates which regime of synchronization we have(from numerical simulation of 1D ddGPE(\ref{1D GP})). There is a well pronounced region for moderate pumps and detunings where the phase derivative is nonzero in a very narrow area. This corresponds to the kinks formation, when phase $\theta$ derivative is noticeably far from 0 only close to the kink kernel. The decreasing of $\sigma$ corresponds to decreasing of the kink size. On the left to kinks region, we observe weak synchronization when the condensate phase growth is almost linear with only slight modification induced by resonant pumping. Green region corresponds to the case when resonant drive is large enough to imprint its own phase to the condensate.}
\end{figure}

The results on the condensate dynamics are summarized in Fig.~\ref{fig3}.
We scanned the range of detunings $\tilde{\delta}$ and range of resonant pump amplitudes $f$, see Fig.~\ref{fig3}, and watched the quantity $\sigma = \sqrt{\int \partial_{\theta} \phi \theta^2 d\theta }/2\pi $ calculated for the stationary solutions. The introduced quantity shows the ratio of the characteristic scale of the phase evolution to the length of the ring. We found that there are three different regimes of the condensate dynamics. In the first regime the coherent pump introduces such a strong perturbation that changes the topological charge of the condensate so that the drive and the condensate become of the same topological charge and get synchronized (green region in Fig.~\ref{fig3}). 

In the second regime, the phase gradient is quasi-uniform and the frequency of the condensate is nearly equal to the frequency of the free-running condensate (condensate is not affected by the coherent pump). In the parameter plane of Fig.~\ref{fig3}, this regime occurs in the area shown by reddish color indicating that the condensate phase gradient distribution is flat. 

The most interesting for us regime is the regime of the phase slip where most of the time the phase of the condensate in a chosen point is locked to the phase of the coherent drive but at some moments of time the phase of the condensate changes by $2\pi$. These periodic phase slips result in the formation of equidistant sidebands around the spectral line corresponding to the frequency of the coherent drive.

As is has already been shown above, in the phase-slip regime the phase of the condensate is a step-like function of the angle moving at a constant velocity.  
Such a behavior leads to the shift of the condensate average frequency $<\Delta\omega> = \Omega = \frac{2\pi}{T}$, where $\Omega$ is the angular velocity of the kink, $T$ is the duration of the kink round trip.
In Fig.~\ref{fig3} this regime corresponds to the area of blueish color showing that the width of the phase step is much less compared to the length of the ring so that the phase step is clearly visible. Let us remark that the width  of the step decreases with increase of the coherent pump strength $f$ until the system switches to a different regime (green area in Fig.~\ref{fig3}).


\subsection{Kinks motion}

\begin{figure}[tb!]
\begin{center}
\includegraphics[width=\linewidth]{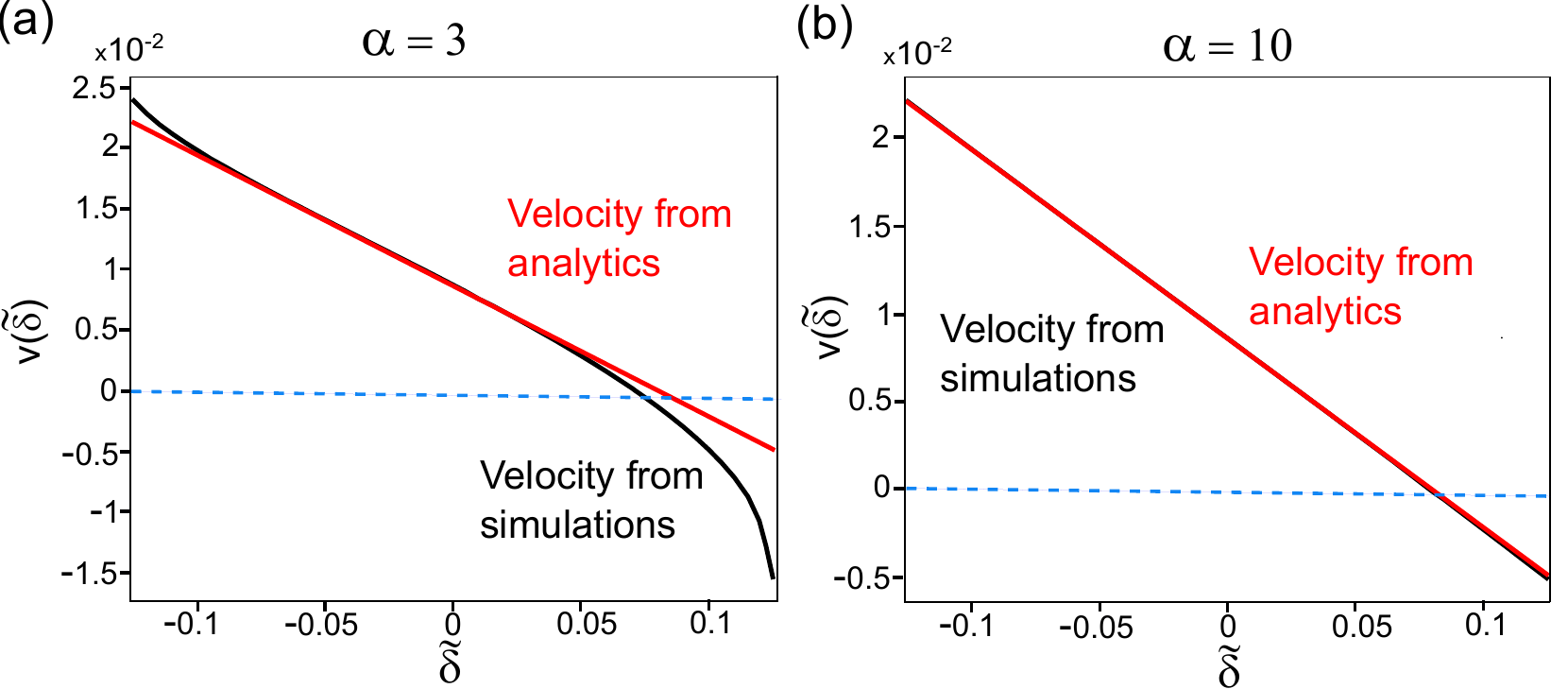}
\end{center}
\caption{ \label{fig4}
(a) and (b) Kink's velocities as functions of detuning $\delta$ for $\alpha = 3$ and $\alpha = 10$. Black lines correspond to numerical simulatons of Eq.(\ref{1D GP}). Red lines correspond to analytical formula (\ref{kinks velocity}). Blue dashed lines corresponds to zero velocity($v=0$).}
\end{figure}

 
In this subsection we will focus on the motion of the kink for various detunings of the resonant drive $\tilde{\delta}$ with respect to the condensate frequency. We perform numerical simulations and plot the dependencies of the kink velocities on the detuning $\tilde \delta$, see  Figure \ref{fig4} showing them for $\alpha=3$ (a) and $\alpha=10$ (b). It is seen that for low velocities this dependency can be approximated well by a straight line given by a perturbation theory developed in the next section.  

\begin{figure}[!t]
\begin{center}
\includegraphics[width=0.9\linewidth]{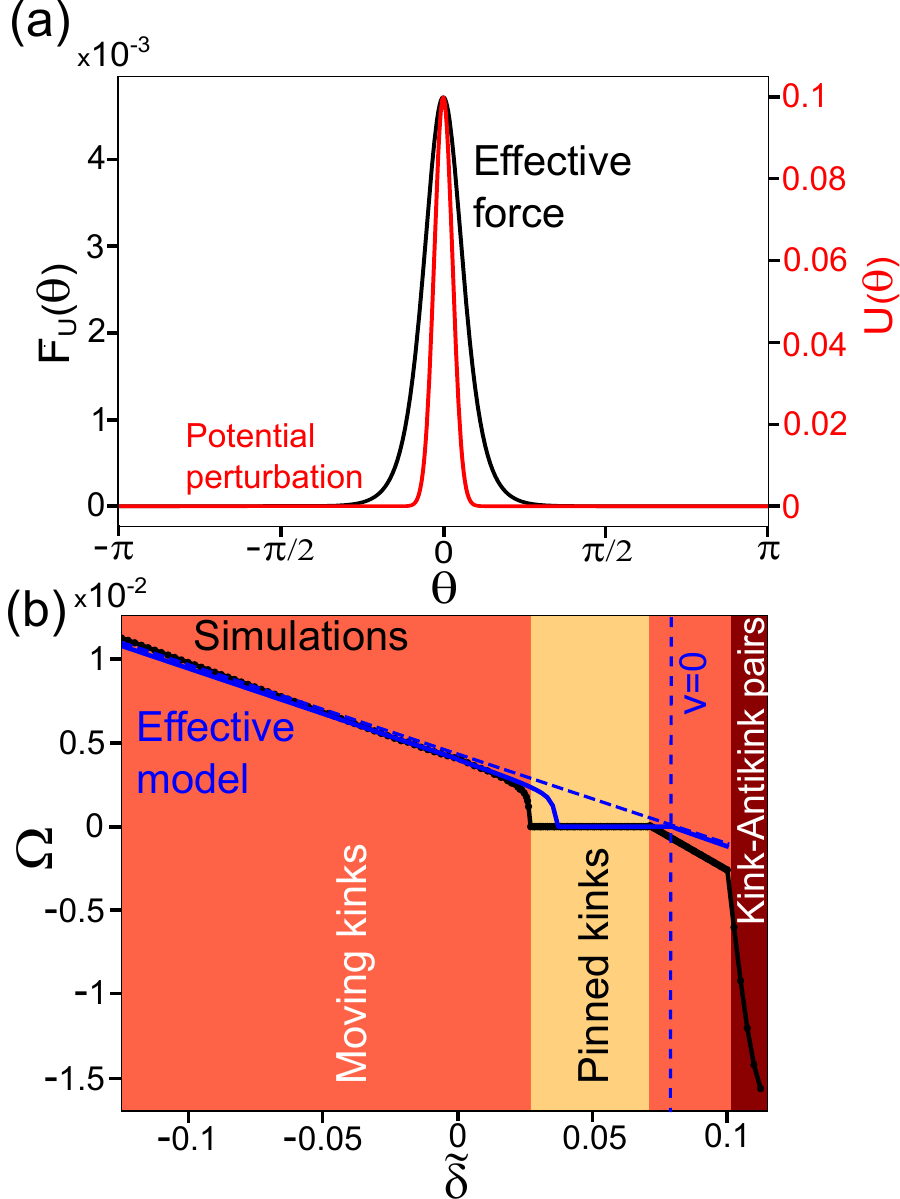}
\end{center}
\caption{ \label{fig5}
(a) The red line shows the angular-dependent part of the effective potential serving as the inhomogeneity trapping the kinks. In the framework of the developed perturbation theory, the action of this potential is accounted in terms of an effective force (\ref{Force from potential}) acting on the kink, see the text. The dependency of this effective force on the angle is shown by the black curve. (b) 
The average kink velocities ( equal to the average phase growth rates ) extracted from the numerical simulations and calculated by the perturbation theory are shown by the black and blue lines correspondingly. 
The blue dashed line shows the analytically found ( Eq.(\ref{kinks velocity}) ) angular velocity of the free moving kink ( in the absence of the potential barrier) as a function of the detuning $\delta$. It is seen that the velocity of the free-running kink changes its sign at some detuning $\tilde{\delta}$. For our parameters the perturbation theory yields $\tilde{\delta} \approx 0.08$. This detuning is marked by the vertical dashed line. Let us note that in the numerical simulations the direction of the motion of the free-running king gets reversed at slightly different value of the detuning $\tilde{\delta} = 0.0711$ which is equal to the border between the areas shown by yellow and light red colors with in precision of our numerical calculations. The numerical simulations revealed that, depending on the detuning, the condensate exhibits different kinds of dynamics: the regime of a single moving kink, the regime of a single kink pinned at the inhomogeneity and the regime where the kink-antikink pair are continually appear and annihilate. The ranges of the detunings where these regimes take place are highlighted by the bright red, yellow and dark red colors correspondingly.}
\end{figure}

Then, we turn to the dynamics of a kink in a presence of localized potential perturbation $U(\theta) = U_0 e^{-(\frac{\theta}{w})^2}$(see Fig.~\ref{fig5} (a)). 
The inhomogeneities of such a kind can be caused by small spatial disorder inevitably present in real microcavities, they can also be created on purpose by the microstructuring of the sample. Spatially non-uniform gain can also produce effective pinning potential for the kinks.  
The inhomogeneities can be accounted by replacing $\delta\rightarrow \delta + U(\theta)$ in the Gross-Pitaevskii equation (\ref{1D GP}). 

We numerically studied the dynamics of the condensate in the presence of a weak localized potential $U$ with the  amplitude $U_0 = 0.1$ and width $w = 0.125$. We observed that, as in the spatially uniform case, at small detunings the dynamics can be understood in terms of the motion of the kinks. The numerically found dependency of the average detuning of the condensate average frequency($<\Delta w> = \Omega$) from the frequency of the coherent pump is shown in Fig.~\ref{fig5}(b).

We can distinguish three different regimes: moving kinks, kinks pinned at the inhomogeneity and, at relatively large detunings, we also observe the dynamics that can be seen as the repeating acts of the kink -- anti-kink pairs creation, their motion,  collisions and annihilation, see Fig.~\ref{fig5}(b) and \textbf{Supplementary movies}. In the simulations discussed here and below we take $\alpha = 3$.

Let us note that only kinks moving with the positive velocity can be stopped by the attractive conservative potential (see Fig.~\ref{fig5}(b) showing the velocity of the kink in the absence of the trapping potential (blue dashed line) as a function of detuning overlapped with the synchronization diagram). The repelling conservative potential can trap only the kinks moving with the negative velocities. In its turn, anti-kinks moving with positive velocities can be trapped only by repelling potential and moving with negative velocities - by attracting potentials only. This effect in a way resembles the diode effect. Below we explain this effect by quasi-particle description of the kink motion.  

Finally, for the sake of completeness, we studied the dynamics in the system with $\Delta n = 2$  where two kinks forms (in general, the number of kinks is simply $N_{kinks} = \Delta n$, negative values corresponds to anti-kinks). Let us remark that to accommodate a large number of kinks we have to work with the annual condensates of bigger radius and this can be complicated from the experimental point of view.   

\begin{figure}[!tb]
\begin{center}
\includegraphics[width=\linewidth]{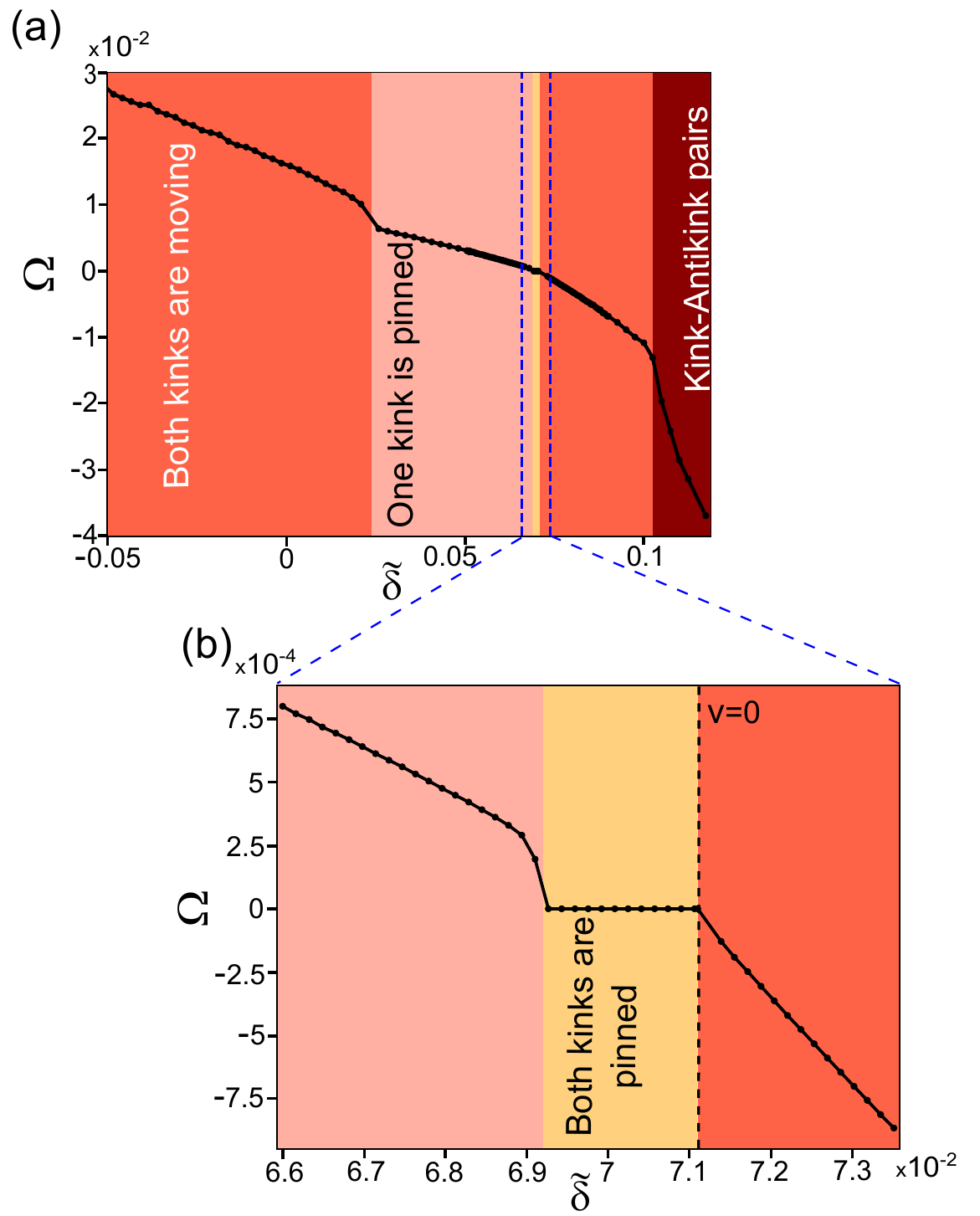}
\end{center}
\caption{ \label{fig6}
(a)The numerically found average phase growth rate $\Omega = \frac{2\pi}{T}$ ($T$ is a roundtrip period for one kink) in the presence of localized potential perturbation is shown by the black line for the case of two kinks formed in the system (topological charges difference is $\Delta n = 2$). The system has several dynamics regimes: both kinks are always in motion, one kink is always pinned while another is in motion, both kinks are pinned and the regime where kink-antikink pairs generation occurs (See \textbf{supplementary movies}) (b) The same as (a) but zoomed in to show the regime where both kinks are pinned. Black dashed line shows the detuning $\tilde{\delta}$ where the numerical kink's velocity is 0.
}
\end{figure}

The results on the dynamics of the kinks are summarized in Fig.~\ref{fig6}. In this case the dynamics becomes richer, in particular there may be the case where both kinks are always moving with nonzero velocities, the case where one kink is moving and the other is pinned on potential (this dynamics is illustrated by the Supplementary movie), the case of two kinks pinned on the potential and the case of repeatedly creating and annihilating kink -- anti-kink pairs.

\section{Quasi-particle description of kinks dynamics}

To understand the dynamics of the kinks, we develop a perturbation theory describing the dynamics of the kink in terms of its velocity. We start with the equation for the phase of the condensate (\ref{effective phase equation}) and consider the limit of large $\alpha$ so that for the small detunings $\delta$ the phase distribution can be well approximated by a sine-Gordon kink solution $\phi_0=4\arctan(e^{m(\theta - \theta_0}) $, $m^2 = f/J$, which is the solution of the equation $J\partial^2_{\theta} \phi+ f\sin \phi=0$. The position of the kink is characterized by the coordinate $\theta_0$ of the kink center (where $|\partial \phi_{\theta}|$ has a maximum). To account for the perturbations we sought the solution in the form $\phi=\phi_0+\chi$ where $\chi$ is a small correction. It can happen that the equation for $\chi$ is such that its solution grows in time and thus is not small at large $t$. To eliminate this problem we use the standard method and allow the coordinate of the kink center $\theta_0$ to be a function of time. Then we can write the condition ensuring that the correction $\chi$ remains small at all times. The condition reads  
\begin{equation}
    \int_{-\pi}^{\pi}\left(\dot{\theta_0} \phi_0' + \tilde{\delta} - J(\phi_0')^2 - f\cos(\phi_0)\right)\phi_0' \,d\theta= 0.
    \label{solvability condition}
\end{equation}

Actually, this solvability condition is nothing else but the orthogonality condition of the right-hand side of the equation for small correction $\chi$ to the eigenfunctions corresponding to zero eigenvalues of the operator standing in the left-hand side of the equation for $\chi$. 

Integrals in the Eq. (\ref{solvability condition}) could be easily computed analytically. Then the equation for the kink's center $\theta_0$ can be written as
\begin{equation}
\dot{\theta_0} = \frac{2\pi m^2 J-\pi\tilde{\delta}}{4m}.
\label{kinks velocity}
\end{equation}
From Eq. (\ref{kinks velocity}) it is possible to see that in the axially symmetric system the kinks move with a constant velocity proportional to the effective detuning of resonant drive from the condensate's frequency $v_0=\frac{2\pi m^2 J-\pi\tilde{\delta}}{4m}$. This fact fits well with the results of numerical simulations discussed in the previous section (see Fig.~\ref{fig4}). 

Let us remark that the kink velocity changes monotonically with $\delta$ and changes its sign at a critical detuning $\tilde{\delta}$, see Fig.~\ref{fig5}. For the parameters we use in our numerical simulations $\tilde{\delta} \approx 0.0711$ and the perturbation theory gives $\tilde{\delta} \approx 0.08$.  
The discrepancy is explained by relatively small value of $\alpha = 3$ (let us remind that the applicability condition for the perturbation theory is $\alpha \gg 1$). For larger values of $\alpha$ the agreement between the numerics and analytics is much better, see Fig.~\ref{fig4} (b) showing the dependencies of the velocity on the detuning for $\alpha=10$. Let us also remark that the agreement is better for small detunings which is also explained by the applicability of the perturbation theory requiring that $\tilde{\delta} = \delta - \alpha\ll1$. The range of detunings $\delta$ shown in Fig.~\ref{fig4} fits this condition and thus one can expect that the analytical results should be in agreement with numerical simulations. For large detunings the kinks get destroyed and thus the polariton dynamics changes drastically, see Fig.~\ref{fig1} (e).


The effect of small perturbations making the detuning of the pump to be spatially nonuniform can also be accounted by the perturbation theory.  To consider potential perturbation we replaced $\delta\rightarrow \delta + U(\theta)$. The solvability condition Eq. (\ref{solvability condition}) in a presence of perturbation has the form 
\begin{equation}
    \int_{-\pi}^{\pi}\left(\dot{\theta_0} \phi_0' + \tilde{\delta} + U(\theta) - J(\phi_0')^2 - f\cos(\phi_0)\right)\phi_0'\,d\theta = 0.
    \label{solvability condition 2}
\end{equation}
The equation for the kinks coordinate is then 
\begin{equation}
    \dot{\theta_0} = v_0 - \operatorname{F}_{U}(\theta_0),
    \label{Dynamics with well}
\end{equation}
where $v_0$ is free kink velocity given by Eq. (\ref{kinks velocity})
\begin{equation}
    \operatorname{F}_{U}(\theta_0) = \frac{1}{8m}\int^{\pi}_{-\pi}U(\theta) \phi_0'(\theta-\theta_0) d\theta
    \label{Force from potential}
\end{equation}

Let us note that, from the formal point of view, Eq. (\ref{Dynamics with well}) is a viscous motion of a particle under the action of the force $v_0 - \operatorname{F}_{U}(\theta_0)$. The force consists of two parts. The first one is associated with the detuning and the second (\ref{Force from potential}) shown in Fig.~\ref{fig5} (a) is associated with the effect of the spatially nonuniform detuning. The equilibrium points (stable and unstable ones) exist if $min(\operatorname{F}_{U}) <v_0 < max(\operatorname{F}_{U})$. 

The developed perturbation theory describes the results of numerical simulations well, see Fig.~\ref{fig5} where the average growth rate $\Omega$ of phase difference between the coherent drive and the condensate are shown. The discrepancy in this case as well as in the case of spatially uniform system  can be explained by relatively small value of $\alpha$.

For the conservative attracting potential $\operatorname{F}_{U}$ is positive. Therefore, this potential can only stop the kinks moving at positive velocity less than $max(\operatorname{F}_{U})$. Correspondingly, the conservative repelling potential produces the force such that $\operatorname{F}_{U}<0 $. Therefore this potential can stop only the kinks moving at negative velocities with absolute value $-min(\operatorname{F}_{U})$. This agrees with the results of numerical simulations of Eq.(\ref{1D GP}), see the corresponding discussion in the previous section. Note that this effect resembles the behavior of charge carriers in diodes: they easily move in one direction only.




Another possibility to provide phase-locking of the condensate to the external coherent pump is to use spatially non-uniform incoherent pump $\gamma = \gamma_0 + \Delta \gamma(\theta)$ which leads to the same dynamics. The phase locking can also be caused by the spatial dependence of the amplitude of the coherent drive  $f = f_0 + \Delta f(\theta)$. The perturbation theory extended to the latter cases is presented in Appendix \textbf{B}. 

Let us remark here that for the deviation of the coherent drive amplitude from its man value described by a Gaussian function the effective force is asymmetric and changes its sign, see Fig.~\ref{fig7}. So in this case, the kinks moving with either positive or negative velocities can be trapped by the inhomogeneity. 


\section{Conclusion}

The studies reported in the paper reveals that the synchronization of ring shaped polariton condensates to the coherent drive depends crucially on the difference of the topological charges of the condensates and the drive. It was shown by numerical simulations of 2D and 1D systems that in axially symmetric system the condensate can be phase-locked to the coherent drive only if they have the same topological charges. In the case of different topological charges the typical dynamics is that the mutual phase at a chosen point stays nearly constant for most of the time by at some moments experienced fast changes of the phase by $2\pi$. This process is periodic and results in the finite difference between the mean frequency of the condensate and the frequency of the drive.

To explain this effect, we considered the case of weak coherent drive and show that then the condensate dynamics can be reduced to the dynamics of the mutual phase between the coherent drive and the condensate. We have derived the equation for the phase and have shown that the equation can have solutions in the form of moving kinks - step-like solution where each step corresponds to the change of the mutual phase by $2\pi$. Every time the kink passes a point the phase slip occurs and the mutual phase between the condensate and the drive changes by $2\pi$. 

We also consider how the axial symmetry breaking affects the dynamics of the kinks and, therefore, the synchronization of the condensate to the drive. It was shown that the kinks can be trapped by the inhomogeneities. The trapping of the kinks stops the periodic phase slips and this way the condensate gets locked to the drive. We considered to trapping of two kinks on one inhomogeneity and demonstrated that there are three different kinds of the behaviors of the kinks. The first regime is when the inhomogeneity is too weak and both kinks are always in motion. The second regime is the strong inhomogeneity that pins both kinks making them resting. In the intermediate case one kink is pinned and the other is moving until it reaches the inhomogeneity. Then the latter kink pushes the former one out of the inhomogeneity setting it in motion and gets pinned on the inhomogeneity. 

To explain the effects observed in numerical simulations we developed a perturbation theory describing the dynamics of the kinks in terms of the motion of its centers coordinates. The perturbative approach shows excellent agreement with the numerics for large values of so-called blue shift insuring the applicability of the theory. However semi-qualitative agreement between the numerical simulations and the perturbation theory can be achieved for experimentally realistic values of the blue shift. The developed theory allows to explain main features of the synchronization including the kinks trapping on the inhomogeneities. It is interesting to note that symmetric potential can produce asymmetric effective force trapping the kinks moving in one direction and not trapping kinks moving in the opposite direction. 

We believe that the reported results can enhance our understanding of the dynamics of polaritons under the action of the external pumps in the axially symmetric systems as well as in the presence of the inhomogeneities. This can be useful from the point of view of the control of the polariton condensates by weak laser beam tuned in the resonance to the condensates, creation of the coherent states of large number of polariton droplets, design of polariton lasers and, potentially, for optical calculations.

\section{Acknowledgments}
This work was supported by Priority 2030 Federal Academic Leadership Program. S.V.K., A.N.O., and O.I.U. acknowledge financial support. from the Institute for Basic Science (IBS) in the Republic of Korea through YSF Project No. IBS-R024-Y3 O.I.U. acknowledges Project No. IBS-R024-D1.

\section{Appendix A. Calculating coefficients for effective 1D Gross-Pitaevskii equation.}

Assuming that we know the radial part of the condensate wavefunction in the trap $F(r)$, coefficients in Eq.(\ref{1D GP dim}) could be calculated via the following expressions $\Gamma = \frac{R^2P_R}{2\hbar\gamma_R\Gamma_R}\int|F(r)|^4rdr$,  $U =  \frac{1}{\hbar}(g_c-\frac{P_R R}{\Gamma_R^2})\int|F(r)|^4rdr$, $J = \frac{\hbar}{2 m_p}\int\frac{1}{r^2}|F(r)|^2rdr\sim \frac{1}{r_0^2}$, where $r_0$ is the trap radius. Here we assume that function $F(r)$ fits normalization condition $\int |F(r)|^2rdr = 1$. 

\section{Appendix B. Dynamics in a presence of a perturbation in coherent pumping}
As it has been already mentioned in the text, to achieve kinks localization we can add localized perturbation to coherent pumping $f = f - \Delta f(\theta)$. The effective dynamics equation in this case is similar to the already observed scenario of potential perturbation

\begin{equation}
    \dot{\theta_0} = v_0 - \operatorname{F}_{\Delta f}(\theta_0),
    \label{Dynamics with resonant perturbation}
\end{equation}
however the expression for the function in the right part is significantly different
\begin{align}
    &\operatorname{F}_{\Delta f}(\theta_0) = \frac{1}{8m}\int^{\pi}_{-\pi}\Delta f(\theta)\big[\cos(\phi_0'(\theta-\theta_0)) - \\
    &\alpha\sin(\phi_0'(\theta-\theta_0))\big] \phi_0'(\theta-\theta_0) d\theta \nonumber
    \label{Force from resonant perturbation}
\end{align}

\begin{figure}[!t]
\begin{center}
\includegraphics[width=\linewidth]{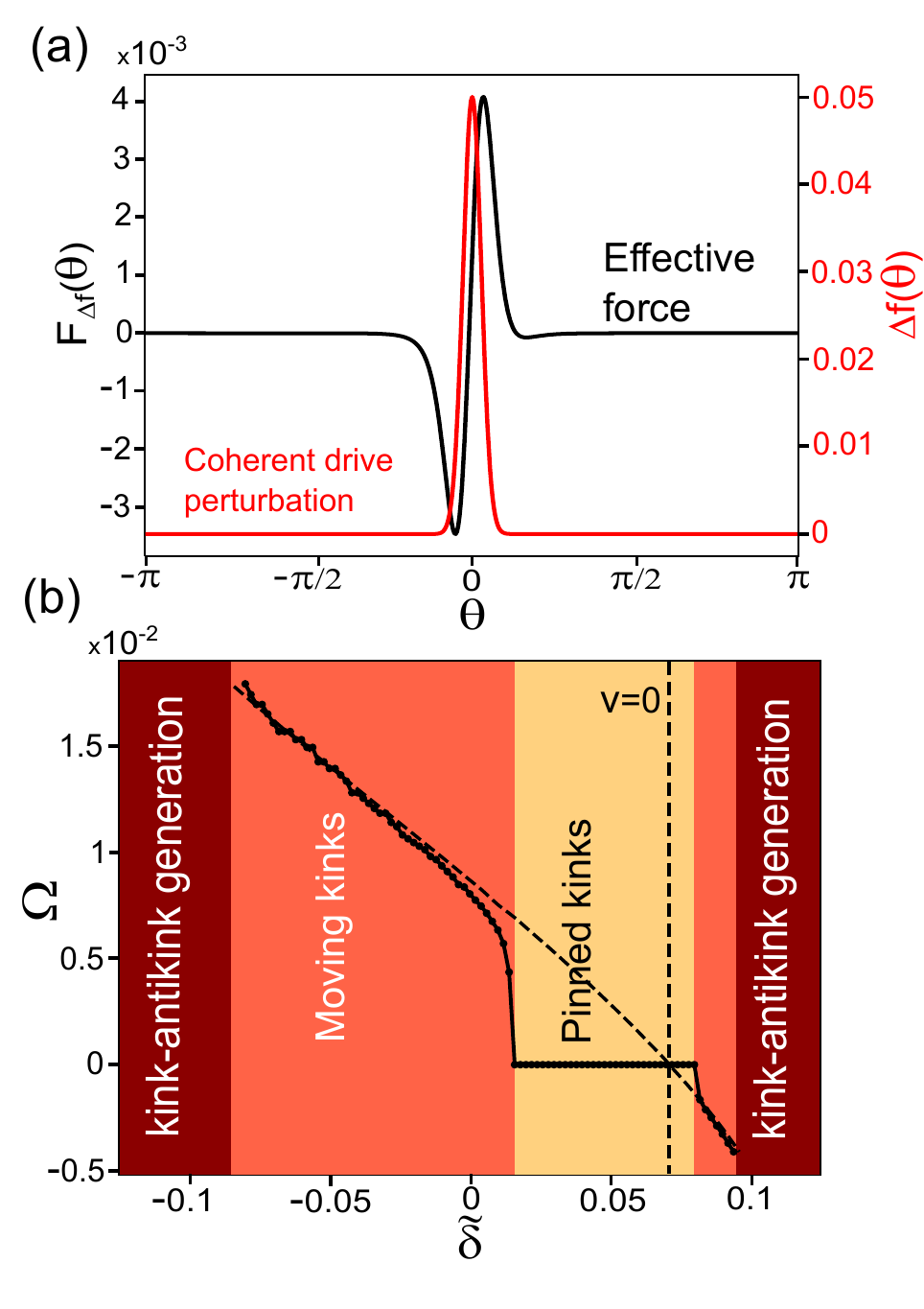}
\end{center}
\caption{ \label{fig7}
(a) Black line - effective force acting on the kink, red line - perturbation in the coherent pumping in the form $\Delta f(\theta) = 0.05 e^{-(\frac{\theta}{0.125})^2}$. (b)Average phase growth rate $\Omega = \frac{2\pi}{T}$ in a presence of localized coherent drive amplitude perturbation derived from direct numerical simulation of 1D GP equation \eqref{1D GP}. Colored sections represents different dynamics regimes for different detunings regions. Black dashed horizontal line represents free kink velocity presented on Fig \ref{fig4}. Black dashed vertical line shows the detuning $\tilde{\delta}$ where the numerical kinks velocity is 0.
}
\end{figure}
As one can see in Fig.~\ref{fig7}, in this case force changes its sign indicating that pinning for kinks of both positive and negative velocities is possible.

\bibliography{APFBibl}

\begin{thebibliography}{46}%
\makeatletter
\providecommand \@ifxundefined [1]{%
 \@ifx{#1\undefined}
}%
\providecommand \@ifnum [1]{%
 \ifnum #1\expandafter \@firstoftwo
 \else \expandafter \@secondoftwo
 \fi
}%
\providecommand \@ifx [1]{%
 \ifx #1\expandafter \@firstoftwo
 \else \expandafter \@secondoftwo
 \fi
}%
\providecommand \natexlab [1]{#1}%
\providecommand \enquote  [1]{``#1''}%
\providecommand \bibnamefont  [1]{#1}%
\providecommand \bibfnamefont [1]{#1}%
\providecommand \citenamefont [1]{#1}%
\providecommand \href@noop [0]{\@secondoftwo}%
\providecommand \href [0]{\begingroup \@sanitize@url \@href}%
\providecommand \@href[1]{\@@startlink{#1}\@@href}%
\providecommand \@@href[1]{\endgroup#1\@@endlink}%
\providecommand \@sanitize@url [0]{\catcode `\\12\catcode `\$12\catcode `\&12\catcode `\#12\catcode `\^12\catcode `\_12\catcode `\%12\relax}%
\providecommand \@@startlink[1]{}%
\providecommand \@@endlink[0]{}%
\providecommand \url  [0]{\begingroup\@sanitize@url \@url }%
\providecommand \@url [1]{\endgroup\@href {#1}{\urlprefix }}%
\providecommand \urlprefix  [0]{URL }%
\providecommand \Eprint [0]{\href }%
\providecommand \doibase [0]{https://doi.org/}%
\providecommand \selectlanguage [0]{\@gobble}%
\providecommand \bibinfo  [0]{\@secondoftwo}%
\providecommand \bibfield  [0]{\@secondoftwo}%
\providecommand \translation [1]{[#1]}%
\providecommand \BibitemOpen [0]{}%
\providecommand \bibitemStop [0]{}%
\providecommand \bibitemNoStop [0]{.\EOS\space}%
\providecommand \EOS [0]{\spacefactor3000\relax}%
\providecommand \BibitemShut  [1]{\csname bibitem#1\endcsname}%
\let\auto@bib@innerbib\@empty
\bibitem [{\citenamefont {Sanvitto}\ and\ \citenamefont {Timofeev}(2012)}]{sanvitto2012exciton}%
  \BibitemOpen
  \bibfield  {author} {\bibinfo {author} {\bibfnamefont {D.}~\bibnamefont {Sanvitto}}\ and\ \bibinfo {author} {\bibfnamefont {V.}~\bibnamefont {Timofeev}},\ }\href@noop {} {\emph {\bibinfo {title} {Exciton Polaritons in Microcavities: New Frontiers}}},\ Vol.\ \bibinfo {volume} {172}\ (\bibinfo  {publisher} {Springer Science \& Business Media},\ \bibinfo {year} {2012})\BibitemShut {NoStop}%
\bibitem [{\citenamefont {Kasprzak}\ \emph {et~al.}(2006)\citenamefont {Kasprzak}, \citenamefont {Richard}, \citenamefont {Kundermann}, \citenamefont {Baas}, \citenamefont {Jeambrun}, \citenamefont {Keeling}, \citenamefont {Marchetti}, \citenamefont {Szyma{\'n}ska}, \citenamefont {Andr{\'e}}, \citenamefont {Staehli} \emph {et~al.}}]{kasprzak2006bose}%
  \BibitemOpen
  \bibfield  {author} {\bibinfo {author} {\bibfnamefont {J.}~\bibnamefont {Kasprzak}}, \bibinfo {author} {\bibfnamefont {M.}~\bibnamefont {Richard}}, \bibinfo {author} {\bibfnamefont {S.}~\bibnamefont {Kundermann}}, \bibinfo {author} {\bibfnamefont {A.}~\bibnamefont {Baas}}, \bibinfo {author} {\bibfnamefont {P.}~\bibnamefont {Jeambrun}}, \bibinfo {author} {\bibfnamefont {J.~M.~J.}\ \bibnamefont {Keeling}}, \bibinfo {author} {\bibfnamefont {F.}~\bibnamefont {Marchetti}}, \bibinfo {author} {\bibfnamefont {M.}~\bibnamefont {Szyma{\'n}ska}}, \bibinfo {author} {\bibfnamefont {R.}~\bibnamefont {Andr{\'e}}}, \bibinfo {author} {\bibfnamefont {J.}~\bibnamefont {Staehli}}, \emph {et~al.},\ }\bibfield  {title} {\bibinfo {title} {Bose--einstein condensation of exciton polaritons},\ }\href@noop {} {\bibfield  {journal} {\bibinfo  {journal} {Nature}\ }\textbf {\bibinfo {volume} {443}},\ \bibinfo {pages} {409} (\bibinfo {year} {2006})}\BibitemShut {NoStop}%
\bibitem [{\citenamefont {Butov}(2007)}]{butov2007polariton}%
  \BibitemOpen
  \bibfield  {author} {\bibinfo {author} {\bibfnamefont {L.~V.}\ \bibnamefont {Butov}},\ }\bibfield  {title} {\bibinfo {title} {A polariton laser},\ }\href@noop {} {\bibfield  {journal} {\bibinfo  {journal} {Nature}\ }\textbf {\bibinfo {volume} {447}},\ \bibinfo {pages} {540} (\bibinfo {year} {2007})}\BibitemShut {NoStop}%
\bibitem [{\citenamefont {Schneider}\ \emph {et~al.}(2013)\citenamefont {Schneider}, \citenamefont {Rahimi-Iman}, \citenamefont {Kim}, \citenamefont {Fischer}, \citenamefont {Savenko}, \citenamefont {Amthor}, \citenamefont {Lermer}, \citenamefont {Wolf}, \citenamefont {Worschech}, \citenamefont {Kulakovskii} \emph {et~al.}}]{schneider2013electrically}%
  \BibitemOpen
  \bibfield  {author} {\bibinfo {author} {\bibfnamefont {C.}~\bibnamefont {Schneider}}, \bibinfo {author} {\bibfnamefont {A.}~\bibnamefont {Rahimi-Iman}}, \bibinfo {author} {\bibfnamefont {N.~Y.}\ \bibnamefont {Kim}}, \bibinfo {author} {\bibfnamefont {J.}~\bibnamefont {Fischer}}, \bibinfo {author} {\bibfnamefont {I.~G.}\ \bibnamefont {Savenko}}, \bibinfo {author} {\bibfnamefont {M.}~\bibnamefont {Amthor}}, \bibinfo {author} {\bibfnamefont {M.}~\bibnamefont {Lermer}}, \bibinfo {author} {\bibfnamefont {A.}~\bibnamefont {Wolf}}, \bibinfo {author} {\bibfnamefont {L.}~\bibnamefont {Worschech}}, \bibinfo {author} {\bibfnamefont {V.~D.}\ \bibnamefont {Kulakovskii}}, \emph {et~al.},\ }\bibfield  {title} {\bibinfo {title} {An electrically pumped polariton laser},\ }\href@noop {} {\bibfield  {journal} {\bibinfo  {journal} {Nature}\ }\textbf {\bibinfo {volume} {497}},\ \bibinfo {pages} {348} (\bibinfo {year} {2013})}\BibitemShut {NoStop}%
\bibitem [{\citenamefont {Nitsche}\ \emph {et~al.}(2014)\citenamefont {Nitsche}, \citenamefont {Kim}, \citenamefont {Roumpos}, \citenamefont {Schneider}, \citenamefont {Kamp}, \citenamefont {H{\"o}fling}, \citenamefont {Forchel},\ and\ \citenamefont {Yamamoto}}]{nitsche2014algebraic}%
  \BibitemOpen
  \bibfield  {author} {\bibinfo {author} {\bibfnamefont {W.~H.}\ \bibnamefont {Nitsche}}, \bibinfo {author} {\bibfnamefont {N.~Y.}\ \bibnamefont {Kim}}, \bibinfo {author} {\bibfnamefont {G.}~\bibnamefont {Roumpos}}, \bibinfo {author} {\bibfnamefont {C.}~\bibnamefont {Schneider}}, \bibinfo {author} {\bibfnamefont {M.}~\bibnamefont {Kamp}}, \bibinfo {author} {\bibfnamefont {S.}~\bibnamefont {H{\"o}fling}}, \bibinfo {author} {\bibfnamefont {A.}~\bibnamefont {Forchel}},\ and\ \bibinfo {author} {\bibfnamefont {Y.}~\bibnamefont {Yamamoto}},\ }\bibfield  {title} {\bibinfo {title} {Algebraic order and the berezinskii-kosterlitz-thouless transition in an exciton-polariton gas},\ }\href@noop {} {\bibfield  {journal} {\bibinfo  {journal} {Physical Review B}\ }\textbf {\bibinfo {volume} {90}},\ \bibinfo {pages} {205430} (\bibinfo {year} {2014})}\BibitemShut {NoStop}%
\bibitem [{\citenamefont {Lovett}\ \emph {et~al.}(2023)\citenamefont {Lovett}, \citenamefont {Walker}, \citenamefont {Osipov}, \citenamefont {Yulin}, \citenamefont {Naik}, \citenamefont {Whittaker}, \citenamefont {Shelykh}, \citenamefont {Skolnick},\ and\ \citenamefont {Krizhanovskii}}]{lovett2023observation}%
  \BibitemOpen
  \bibfield  {author} {\bibinfo {author} {\bibfnamefont {S.}~\bibnamefont {Lovett}}, \bibinfo {author} {\bibfnamefont {P.~M.}\ \bibnamefont {Walker}}, \bibinfo {author} {\bibfnamefont {A.}~\bibnamefont {Osipov}}, \bibinfo {author} {\bibfnamefont {A.}~\bibnamefont {Yulin}}, \bibinfo {author} {\bibfnamefont {P.~U.}\ \bibnamefont {Naik}}, \bibinfo {author} {\bibfnamefont {C.~E.}\ \bibnamefont {Whittaker}}, \bibinfo {author} {\bibfnamefont {I.~A.}\ \bibnamefont {Shelykh}}, \bibinfo {author} {\bibfnamefont {M.~S.}\ \bibnamefont {Skolnick}},\ and\ \bibinfo {author} {\bibfnamefont {D.~N.}\ \bibnamefont {Krizhanovskii}},\ }\bibfield  {title} {\bibinfo {title} {Observation of zitterbewegung in photonic microcavities},\ }\href@noop {} {\bibfield  {journal} {\bibinfo  {journal} {Light: Science \& Applications}\ }\textbf {\bibinfo {volume} {12}},\ \bibinfo {pages} {126} (\bibinfo {year} {2023})}\BibitemShut {NoStop}%
\bibitem [{\citenamefont {Liew}\ \emph {et~al.}(2011)\citenamefont {Liew}, \citenamefont {Shelykh},\ and\ \citenamefont {Malpuech}}]{liew2011polaritonic}%
  \BibitemOpen
  \bibfield  {author} {\bibinfo {author} {\bibfnamefont {T.}~\bibnamefont {Liew}}, \bibinfo {author} {\bibfnamefont {I.}~\bibnamefont {Shelykh}},\ and\ \bibinfo {author} {\bibfnamefont {G.}~\bibnamefont {Malpuech}},\ }\bibfield  {title} {\bibinfo {title} {Polaritonic devices},\ }\href@noop {} {\bibfield  {journal} {\bibinfo  {journal} {Physica E: Low-dimensional Systems and Nanostructures}\ }\textbf {\bibinfo {volume} {43}},\ \bibinfo {pages} {1543} (\bibinfo {year} {2011})}\BibitemShut {NoStop}%
\bibitem [{\citenamefont {Sanvitto}\ and\ \citenamefont {K{\'e}na-Cohen}(2016)}]{sanvitto2016road}%
  \BibitemOpen
  \bibfield  {author} {\bibinfo {author} {\bibfnamefont {D.}~\bibnamefont {Sanvitto}}\ and\ \bibinfo {author} {\bibfnamefont {S.}~\bibnamefont {K{\'e}na-Cohen}},\ }\bibfield  {title} {\bibinfo {title} {The road towards polaritonic devices},\ }\href@noop {} {\bibfield  {journal} {\bibinfo  {journal} {Nature materials}\ }\textbf {\bibinfo {volume} {15}},\ \bibinfo {pages} {1061} (\bibinfo {year} {2016})}\BibitemShut {NoStop}%
\bibitem [{\citenamefont {Berloff}\ \emph {et~al.}(2017)\citenamefont {Berloff}, \citenamefont {Silva}, \citenamefont {Kalinin}, \citenamefont {Askitopoulos}, \citenamefont {T{\"o}pfer}, \citenamefont {Cilibrizzi}, \citenamefont {Langbein},\ and\ \citenamefont {Lagoudakis}}]{berloff2017realizing}%
  \BibitemOpen
  \bibfield  {author} {\bibinfo {author} {\bibfnamefont {N.~G.}\ \bibnamefont {Berloff}}, \bibinfo {author} {\bibfnamefont {M.}~\bibnamefont {Silva}}, \bibinfo {author} {\bibfnamefont {K.}~\bibnamefont {Kalinin}}, \bibinfo {author} {\bibfnamefont {A.}~\bibnamefont {Askitopoulos}}, \bibinfo {author} {\bibfnamefont {J.~D.}\ \bibnamefont {T{\"o}pfer}}, \bibinfo {author} {\bibfnamefont {P.}~\bibnamefont {Cilibrizzi}}, \bibinfo {author} {\bibfnamefont {W.}~\bibnamefont {Langbein}},\ and\ \bibinfo {author} {\bibfnamefont {P.~G.}\ \bibnamefont {Lagoudakis}},\ }\bibfield  {title} {\bibinfo {title} {Realizing the classical xy hamiltonian in polariton simulators},\ }\href@noop {} {\bibfield  {journal} {\bibinfo  {journal} {Nature materials}\ }\textbf {\bibinfo {volume} {16}},\ \bibinfo {pages} {1120} (\bibinfo {year} {2017})}\BibitemShut {NoStop}%
\bibitem [{\citenamefont {Lagoudakis}\ and\ \citenamefont {Berloff}(2017)}]{lagoudakis2017polariton}%
  \BibitemOpen
  \bibfield  {author} {\bibinfo {author} {\bibfnamefont {P.~G.}\ \bibnamefont {Lagoudakis}}\ and\ \bibinfo {author} {\bibfnamefont {N.~G.}\ \bibnamefont {Berloff}},\ }\bibfield  {title} {\bibinfo {title} {A polariton graph simulator},\ }\href@noop {} {\bibfield  {journal} {\bibinfo  {journal} {New Journal of Physics}\ }\textbf {\bibinfo {volume} {19}},\ \bibinfo {pages} {125008} (\bibinfo {year} {2017})}\BibitemShut {NoStop}%
\bibitem [{\citenamefont {Zasedatelev}\ \emph {et~al.}(2019)\citenamefont {Zasedatelev}, \citenamefont {Baranikov}, \citenamefont {Urbonas}, \citenamefont {Scafirimuto}, \citenamefont {Scherf}, \citenamefont {St{\"o}ferle}, \citenamefont {Mahrt},\ and\ \citenamefont {Lagoudakis}}]{zasedatelev2019room}%
  \BibitemOpen
  \bibfield  {author} {\bibinfo {author} {\bibfnamefont {A.~V.}\ \bibnamefont {Zasedatelev}}, \bibinfo {author} {\bibfnamefont {A.~V.}\ \bibnamefont {Baranikov}}, \bibinfo {author} {\bibfnamefont {D.}~\bibnamefont {Urbonas}}, \bibinfo {author} {\bibfnamefont {F.}~\bibnamefont {Scafirimuto}}, \bibinfo {author} {\bibfnamefont {U.}~\bibnamefont {Scherf}}, \bibinfo {author} {\bibfnamefont {T.}~\bibnamefont {St{\"o}ferle}}, \bibinfo {author} {\bibfnamefont {R.~F.}\ \bibnamefont {Mahrt}},\ and\ \bibinfo {author} {\bibfnamefont {P.~G.}\ \bibnamefont {Lagoudakis}},\ }\bibfield  {title} {\bibinfo {title} {A room-temperature organic polariton transistor},\ }\href@noop {} {\bibfield  {journal} {\bibinfo  {journal} {Nature Photonics}\ }\textbf {\bibinfo {volume} {13}},\ \bibinfo {pages} {378} (\bibinfo {year} {2019})}\BibitemShut {NoStop}%
\bibitem [{\citenamefont {Harrison}\ \emph {et~al.}(2022)\citenamefont {Harrison}, \citenamefont {Sigurdsson}, \citenamefont {Alyatkin}, \citenamefont {T{\"o}pfer},\ and\ \citenamefont {Lagoudakis}}]{harrison2022solving}%
  \BibitemOpen
  \bibfield  {author} {\bibinfo {author} {\bibfnamefont {S.~L.}\ \bibnamefont {Harrison}}, \bibinfo {author} {\bibfnamefont {H.}~\bibnamefont {Sigurdsson}}, \bibinfo {author} {\bibfnamefont {S.}~\bibnamefont {Alyatkin}}, \bibinfo {author} {\bibfnamefont {J.~D.}\ \bibnamefont {T{\"o}pfer}},\ and\ \bibinfo {author} {\bibfnamefont {P.~G.}\ \bibnamefont {Lagoudakis}},\ }\bibfield  {title} {\bibinfo {title} {Solving the max-3-cut problem with coherent networks},\ }\href@noop {} {\bibfield  {journal} {\bibinfo  {journal} {Physical Review Applied}\ }\textbf {\bibinfo {volume} {17}},\ \bibinfo {pages} {024063} (\bibinfo {year} {2022})}\BibitemShut {NoStop}%
\bibitem [{\citenamefont {Barrat}\ \emph {et~al.}(2023)\citenamefont {Barrat}, \citenamefont {Tzortzakakis}, \citenamefont {Niu}, \citenamefont {Zhou}, \citenamefont {Paschos}, \citenamefont {Petrosyan},\ and\ \citenamefont {Savvidis}}]{barrat2023superfluid}%
  \BibitemOpen
  \bibfield  {author} {\bibinfo {author} {\bibfnamefont {J.}~\bibnamefont {Barrat}}, \bibinfo {author} {\bibfnamefont {A.}~\bibnamefont {Tzortzakakis}}, \bibinfo {author} {\bibfnamefont {M.}~\bibnamefont {Niu}}, \bibinfo {author} {\bibfnamefont {X.}~\bibnamefont {Zhou}}, \bibinfo {author} {\bibfnamefont {G.}~\bibnamefont {Paschos}}, \bibinfo {author} {\bibfnamefont {D.}~\bibnamefont {Petrosyan}},\ and\ \bibinfo {author} {\bibfnamefont {P.}~\bibnamefont {Savvidis}},\ }\bibfield  {title} {\bibinfo {title} {Superfluid polaritonic qubit in an annular trap},\ }\href@noop {} {\bibfield  {journal} {\bibinfo  {journal} {arXiv preprint arXiv:2308.05555}\ } (\bibinfo {year} {2023})}\BibitemShut {NoStop}%
\bibitem [{\citenamefont {Kalinin}\ and\ \citenamefont {Berloff}(2018)}]{kalinin2018simulating}%
  \BibitemOpen
  \bibfield  {author} {\bibinfo {author} {\bibfnamefont {K.~P.}\ \bibnamefont {Kalinin}}\ and\ \bibinfo {author} {\bibfnamefont {N.~G.}\ \bibnamefont {Berloff}},\ }\bibfield  {title} {\bibinfo {title} {Simulating ising and n-state planar potts models and external fields with nonequilibrium condensates},\ }\href@noop {} {\bibfield  {journal} {\bibinfo  {journal} {Physical review letters}\ }\textbf {\bibinfo {volume} {121}},\ \bibinfo {pages} {235302} (\bibinfo {year} {2018})}\BibitemShut {NoStop}%
\bibitem [{\citenamefont {Chestnov}\ \emph {et~al.}(2019)\citenamefont {Chestnov}, \citenamefont {Kavokin},\ and\ \citenamefont {Yulin}}]{chestnov2019optical}%
  \BibitemOpen
  \bibfield  {author} {\bibinfo {author} {\bibfnamefont {I.}~\bibnamefont {Chestnov}}, \bibinfo {author} {\bibfnamefont {A.}~\bibnamefont {Kavokin}},\ and\ \bibinfo {author} {\bibfnamefont {A.}~\bibnamefont {Yulin}},\ }\bibfield  {title} {\bibinfo {title} {The optical control of phase locking of polariton condensates},\ }\href@noop {} {\bibfield  {journal} {\bibinfo  {journal} {New Journal of Physics}\ }\textbf {\bibinfo {volume} {21}},\ \bibinfo {pages} {113009} (\bibinfo {year} {2019})}\BibitemShut {NoStop}%
\bibitem [{\citenamefont {Ohadi}\ \emph {et~al.}(2016)\citenamefont {Ohadi}, \citenamefont {Gregory}, \citenamefont {Freegarde}, \citenamefont {Rubo}, \citenamefont {Kavokin}, \citenamefont {Berloff},\ and\ \citenamefont {Lagoudakis}}]{ohadi2016nontrivial}%
  \BibitemOpen
  \bibfield  {author} {\bibinfo {author} {\bibfnamefont {H.}~\bibnamefont {Ohadi}}, \bibinfo {author} {\bibfnamefont {R.}~\bibnamefont {Gregory}}, \bibinfo {author} {\bibfnamefont {T.}~\bibnamefont {Freegarde}}, \bibinfo {author} {\bibfnamefont {Y.}~\bibnamefont {Rubo}}, \bibinfo {author} {\bibfnamefont {A.}~\bibnamefont {Kavokin}}, \bibinfo {author} {\bibfnamefont {N.}~\bibnamefont {Berloff}},\ and\ \bibinfo {author} {\bibfnamefont {P.}~\bibnamefont {Lagoudakis}},\ }\bibfield  {title} {\bibinfo {title} {Nontrivial phase coupling in polariton multiplets},\ }\href@noop {} {\bibfield  {journal} {\bibinfo  {journal} {Physical Review X}\ }\textbf {\bibinfo {volume} {6}},\ \bibinfo {pages} {031032} (\bibinfo {year} {2016})}\BibitemShut {NoStop}%
\bibitem [{\citenamefont {Abbarchi}\ \emph {et~al.}(2013)\citenamefont {Abbarchi}, \citenamefont {Amo}, \citenamefont {Sala}, \citenamefont {Solnyshkov}, \citenamefont {Flayac}, \citenamefont {Ferrier}, \citenamefont {Sagnes}, \citenamefont {Galopin}, \citenamefont {Lema{\^\i}tre}, \citenamefont {Malpuech} \emph {et~al.}}]{abbarchi2013macroscopic}%
  \BibitemOpen
  \bibfield  {author} {\bibinfo {author} {\bibfnamefont {M.}~\bibnamefont {Abbarchi}}, \bibinfo {author} {\bibfnamefont {A.}~\bibnamefont {Amo}}, \bibinfo {author} {\bibfnamefont {V.}~\bibnamefont {Sala}}, \bibinfo {author} {\bibfnamefont {D.}~\bibnamefont {Solnyshkov}}, \bibinfo {author} {\bibfnamefont {H.}~\bibnamefont {Flayac}}, \bibinfo {author} {\bibfnamefont {L.}~\bibnamefont {Ferrier}}, \bibinfo {author} {\bibfnamefont {I.}~\bibnamefont {Sagnes}}, \bibinfo {author} {\bibfnamefont {E.}~\bibnamefont {Galopin}}, \bibinfo {author} {\bibfnamefont {A.}~\bibnamefont {Lema{\^\i}tre}}, \bibinfo {author} {\bibfnamefont {G.}~\bibnamefont {Malpuech}}, \emph {et~al.},\ }\bibfield  {title} {\bibinfo {title} {Macroscopic quantum self-trapping and josephson oscillations of exciton polaritons},\ }\href@noop {} {\bibfield  {journal} {\bibinfo  {journal} {Nature Physics}\ }\textbf {\bibinfo {volume} {9}},\ \bibinfo {pages} {275} (\bibinfo {year} {2013})}\BibitemShut {NoStop}%
\bibitem [{\citenamefont {T{\"o}pfer}\ \emph {et~al.}(2020)\citenamefont {T{\"o}pfer}, \citenamefont {Sigurdsson}, \citenamefont {Pickup},\ and\ \citenamefont {Lagoudakis}}]{topfer2020time}%
  \BibitemOpen
  \bibfield  {author} {\bibinfo {author} {\bibfnamefont {J.~D.}\ \bibnamefont {T{\"o}pfer}}, \bibinfo {author} {\bibfnamefont {H.}~\bibnamefont {Sigurdsson}}, \bibinfo {author} {\bibfnamefont {L.}~\bibnamefont {Pickup}},\ and\ \bibinfo {author} {\bibfnamefont {P.~G.}\ \bibnamefont {Lagoudakis}},\ }\bibfield  {title} {\bibinfo {title} {Time-delay polaritonics},\ }\href@noop {} {\bibfield  {journal} {\bibinfo  {journal} {Communications Physics}\ }\textbf {\bibinfo {volume} {3}},\ \bibinfo {pages} {2} (\bibinfo {year} {2020})}\BibitemShut {NoStop}%
\bibitem [{\citenamefont {Pismen}(1999)}]{pismen1999vortices}%
  \BibitemOpen
  \bibfield  {author} {\bibinfo {author} {\bibfnamefont {L.~M.}\ \bibnamefont {Pismen}},\ }\href@noop {} {\emph {\bibinfo {title} {Vortices in nonlinear fields: from liquid crystals to superfluids, from non-equilibrium patterns to cosmic strings}}},\ Vol.\ \bibinfo {volume} {100}\ (\bibinfo  {publisher} {Oxford University Press},\ \bibinfo {year} {1999})\BibitemShut {NoStop}%
\bibitem [{\citenamefont {Dall}\ \emph {et~al.}(2014)\citenamefont {Dall}, \citenamefont {Fraser}, \citenamefont {Desyatnikov}, \citenamefont {Li}, \citenamefont {Brodbeck}, \citenamefont {Kamp}, \citenamefont {Schneider}, \citenamefont {H{\"o}fling},\ and\ \citenamefont {Ostrovskaya}}]{dall2014creation}%
  \BibitemOpen
  \bibfield  {author} {\bibinfo {author} {\bibfnamefont {R.}~\bibnamefont {Dall}}, \bibinfo {author} {\bibfnamefont {M.~D.}\ \bibnamefont {Fraser}}, \bibinfo {author} {\bibfnamefont {A.~S.}\ \bibnamefont {Desyatnikov}}, \bibinfo {author} {\bibfnamefont {G.}~\bibnamefont {Li}}, \bibinfo {author} {\bibfnamefont {S.}~\bibnamefont {Brodbeck}}, \bibinfo {author} {\bibfnamefont {M.}~\bibnamefont {Kamp}}, \bibinfo {author} {\bibfnamefont {C.}~\bibnamefont {Schneider}}, \bibinfo {author} {\bibfnamefont {S.}~\bibnamefont {H{\"o}fling}},\ and\ \bibinfo {author} {\bibfnamefont {E.~A.}\ \bibnamefont {Ostrovskaya}},\ }\bibfield  {title} {\bibinfo {title} {Creation of orbital angular momentum states with chiral polaritonic lenses},\ }\href@noop {} {\bibfield  {journal} {\bibinfo  {journal} {Physical review letters}\ }\textbf {\bibinfo {volume} {113}},\ \bibinfo {pages} {200404} (\bibinfo {year} {2014})}\BibitemShut {NoStop}%
\bibitem [{\citenamefont {Dreismann}\ \emph {et~al.}(2014)\citenamefont {Dreismann}, \citenamefont {Cristofolini}, \citenamefont {Balili}, \citenamefont {Christmann}, \citenamefont {Pinsker}, \citenamefont {Berloff}, \citenamefont {Hatzopoulos}, \citenamefont {Savvidis},\ and\ \citenamefont {Baumberg}}]{dreismann2014coupled}%
  \BibitemOpen
  \bibfield  {author} {\bibinfo {author} {\bibfnamefont {A.}~\bibnamefont {Dreismann}}, \bibinfo {author} {\bibfnamefont {P.}~\bibnamefont {Cristofolini}}, \bibinfo {author} {\bibfnamefont {R.}~\bibnamefont {Balili}}, \bibinfo {author} {\bibfnamefont {G.}~\bibnamefont {Christmann}}, \bibinfo {author} {\bibfnamefont {F.}~\bibnamefont {Pinsker}}, \bibinfo {author} {\bibfnamefont {N.~G.}\ \bibnamefont {Berloff}}, \bibinfo {author} {\bibfnamefont {Z.}~\bibnamefont {Hatzopoulos}}, \bibinfo {author} {\bibfnamefont {P.~G.}\ \bibnamefont {Savvidis}},\ and\ \bibinfo {author} {\bibfnamefont {J.~J.}\ \bibnamefont {Baumberg}},\ }\bibfield  {title} {\bibinfo {title} {Coupled counterrotating polariton condensates in optically defined annular potentials},\ }\href@noop {} {\bibfield  {journal} {\bibinfo  {journal} {Proceedings of the National Academy of Sciences}\ }\textbf {\bibinfo {volume} {111}},\ \bibinfo {pages} {8770} (\bibinfo {year} {2014})}\BibitemShut {NoStop}%
\bibitem [{\citenamefont {Sigurdsson}\ \emph {et~al.}(2014)\citenamefont {Sigurdsson}, \citenamefont {Egorov}, \citenamefont {Ma}, \citenamefont {Shelykh},\ and\ \citenamefont {Liew}}]{sigurdsson2014information}%
  \BibitemOpen
  \bibfield  {author} {\bibinfo {author} {\bibfnamefont {H.}~\bibnamefont {Sigurdsson}}, \bibinfo {author} {\bibfnamefont {O.}~\bibnamefont {Egorov}}, \bibinfo {author} {\bibfnamefont {X.}~\bibnamefont {Ma}}, \bibinfo {author} {\bibfnamefont {I.~A.}\ \bibnamefont {Shelykh}},\ and\ \bibinfo {author} {\bibfnamefont {T.~C.~H.}\ \bibnamefont {Liew}},\ }\bibfield  {title} {\bibinfo {title} {Information processing with topologically protected vortex memories in exciton-polariton condensates},\ }\href@noop {} {\bibfield  {journal} {\bibinfo  {journal} {Physical Review B}\ }\textbf {\bibinfo {volume} {90}},\ \bibinfo {pages} {014504} (\bibinfo {year} {2014})}\BibitemShut {NoStop}%
\bibitem [{\citenamefont {Ma}\ \emph {et~al.}(2016)\citenamefont {Ma}, \citenamefont {Peschel},\ and\ \citenamefont {Egorov}}]{ma2016incoherent}%
  \BibitemOpen
  \bibfield  {author} {\bibinfo {author} {\bibfnamefont {X.}~\bibnamefont {Ma}}, \bibinfo {author} {\bibfnamefont {U.}~\bibnamefont {Peschel}},\ and\ \bibinfo {author} {\bibfnamefont {O.~A.}\ \bibnamefont {Egorov}},\ }\bibfield  {title} {\bibinfo {title} {Incoherent control of topological charges in nonequilibrium polariton condensates},\ }\href@noop {} {\bibfield  {journal} {\bibinfo  {journal} {Physical Review B}\ }\textbf {\bibinfo {volume} {93}},\ \bibinfo {pages} {035315} (\bibinfo {year} {2016})}\BibitemShut {NoStop}%
\bibitem [{\citenamefont {Ma}\ \emph {et~al.}(2020)\citenamefont {Ma}, \citenamefont {Berger}, \citenamefont {A{\ss}mann}, \citenamefont {Driben}, \citenamefont {Meier}, \citenamefont {Schneider}, \citenamefont {H{\"o}fling},\ and\ \citenamefont {Schumacher}}]{ma2020realization}%
  \BibitemOpen
  \bibfield  {author} {\bibinfo {author} {\bibfnamefont {X.}~\bibnamefont {Ma}}, \bibinfo {author} {\bibfnamefont {B.}~\bibnamefont {Berger}}, \bibinfo {author} {\bibfnamefont {M.}~\bibnamefont {A{\ss}mann}}, \bibinfo {author} {\bibfnamefont {R.}~\bibnamefont {Driben}}, \bibinfo {author} {\bibfnamefont {T.}~\bibnamefont {Meier}}, \bibinfo {author} {\bibfnamefont {C.}~\bibnamefont {Schneider}}, \bibinfo {author} {\bibfnamefont {S.}~\bibnamefont {H{\"o}fling}},\ and\ \bibinfo {author} {\bibfnamefont {S.}~\bibnamefont {Schumacher}},\ }\bibfield  {title} {\bibinfo {title} {Realization of all-optical vortex switching in exciton-polariton condensates},\ }\href@noop {} {\bibfield  {journal} {\bibinfo  {journal} {Nature communications}\ }\textbf {\bibinfo {volume} {11}},\ \bibinfo {pages} {897} (\bibinfo {year} {2020})}\BibitemShut {NoStop}%
\bibitem [{\citenamefont {Gnusov}\ \emph {et~al.}(2023)\citenamefont {Gnusov}, \citenamefont {Harrison}, \citenamefont {Alyatkin}, \citenamefont {Sitnik}, \citenamefont {T{\"o}pfer}, \citenamefont {Sigurdsson},\ and\ \citenamefont {Lagoudakis}}]{gnusov2023quantum}%
  \BibitemOpen
  \bibfield  {author} {\bibinfo {author} {\bibfnamefont {I.}~\bibnamefont {Gnusov}}, \bibinfo {author} {\bibfnamefont {S.}~\bibnamefont {Harrison}}, \bibinfo {author} {\bibfnamefont {S.}~\bibnamefont {Alyatkin}}, \bibinfo {author} {\bibfnamefont {K.}~\bibnamefont {Sitnik}}, \bibinfo {author} {\bibfnamefont {J.}~\bibnamefont {T{\"o}pfer}}, \bibinfo {author} {\bibfnamefont {H.}~\bibnamefont {Sigurdsson}},\ and\ \bibinfo {author} {\bibfnamefont {P.}~\bibnamefont {Lagoudakis}},\ }\bibfield  {title} {\bibinfo {title} {Quantum vortex formation in the “rotating bucket” experiment with polariton condensates},\ }\href@noop {} {\bibfield  {journal} {\bibinfo  {journal} {Science Advances}\ }\textbf {\bibinfo {volume} {9}},\ \bibinfo {pages} {eadd1299} (\bibinfo {year} {2023})}\BibitemShut {NoStop}%
\bibitem [{\citenamefont {Choi}\ \emph {et~al.}(2022)\citenamefont {Choi}, \citenamefont {Park}, \citenamefont {Oh}, \citenamefont {Kwon}, \citenamefont {Park}, \citenamefont {Kang}, \citenamefont {Song}, \citenamefont {Ko}, \citenamefont {Sun}, \citenamefont {Savenko}, \citenamefont {Cho},\ and\ \citenamefont {Choi}}]{PhysRevB.105.L060502}%
  \BibitemOpen
  \bibfield  {author} {\bibinfo {author} {\bibfnamefont {D.}~\bibnamefont {Choi}}, \bibinfo {author} {\bibfnamefont {M.}~\bibnamefont {Park}}, \bibinfo {author} {\bibfnamefont {B.~Y.}\ \bibnamefont {Oh}}, \bibinfo {author} {\bibfnamefont {M.-S.}\ \bibnamefont {Kwon}}, \bibinfo {author} {\bibfnamefont {S.~I.}\ \bibnamefont {Park}}, \bibinfo {author} {\bibfnamefont {S.}~\bibnamefont {Kang}}, \bibinfo {author} {\bibfnamefont {J.~D.}\ \bibnamefont {Song}}, \bibinfo {author} {\bibfnamefont {D.}~\bibnamefont {Ko}}, \bibinfo {author} {\bibfnamefont {M.}~\bibnamefont {Sun}}, \bibinfo {author} {\bibfnamefont {I.~G.}\ \bibnamefont {Savenko}}, \bibinfo {author} {\bibfnamefont {Y.-H.}\ \bibnamefont {Cho}},\ and\ \bibinfo {author} {\bibfnamefont {H.}~\bibnamefont {Choi}},\ }\bibfield  {title} {\bibinfo {title} {Observation of a single quantized vortex vanishment in exciton-polariton superfluids},\ }\href {https://doi.org/10.1103/PhysRevB.105.L060502} {\bibfield  {journal} {\bibinfo  {journal} {Phys. Rev. B}\ }\textbf
  {\bibinfo {volume} {105}},\ \bibinfo {pages} {L060502} (\bibinfo {year} {2022})}\BibitemShut {NoStop}%
\bibitem [{\citenamefont {Keeling}\ and\ \citenamefont {Berloff}(2008)}]{keeling2008spontaneous}%
  \BibitemOpen
  \bibfield  {author} {\bibinfo {author} {\bibfnamefont {J.}~\bibnamefont {Keeling}}\ and\ \bibinfo {author} {\bibfnamefont {N.~G.}\ \bibnamefont {Berloff}},\ }\bibfield  {title} {\bibinfo {title} {Spontaneous rotating vortex lattices in a pumped decaying condensate},\ }\href@noop {} {\bibfield  {journal} {\bibinfo  {journal} {Physical review letters}\ }\textbf {\bibinfo {volume} {100}},\ \bibinfo {pages} {250401} (\bibinfo {year} {2008})}\BibitemShut {NoStop}%
\bibitem [{\citenamefont {Chestnov}\ \emph {et~al.}(2021)\citenamefont {Chestnov}, \citenamefont {Yulin}, \citenamefont {Shelykh},\ and\ \citenamefont {Kavokin}}]{chestnov2021dissipative}%
  \BibitemOpen
  \bibfield  {author} {\bibinfo {author} {\bibfnamefont {I.}~\bibnamefont {Chestnov}}, \bibinfo {author} {\bibfnamefont {A.}~\bibnamefont {Yulin}}, \bibinfo {author} {\bibfnamefont {I.}~\bibnamefont {Shelykh}},\ and\ \bibinfo {author} {\bibfnamefont {A.}~\bibnamefont {Kavokin}},\ }\bibfield  {title} {\bibinfo {title} {Dissipative josephson vortices in annular polariton fluids},\ }\href@noop {} {\bibfield  {journal} {\bibinfo  {journal} {Physical Review B}\ }\textbf {\bibinfo {volume} {104}},\ \bibinfo {pages} {165305} (\bibinfo {year} {2021})}\BibitemShut {NoStop}%
\bibitem [{\citenamefont {Harrison}\ \emph {et~al.}(2023)\citenamefont {Harrison}, \citenamefont {Nalitov}, \citenamefont {Lagoudakis},\ and\ \citenamefont {Sigurdsson}}]{harrison2023polariton}%
  \BibitemOpen
  \bibfield  {author} {\bibinfo {author} {\bibfnamefont {S.~L.}\ \bibnamefont {Harrison}}, \bibinfo {author} {\bibfnamefont {A.}~\bibnamefont {Nalitov}}, \bibinfo {author} {\bibfnamefont {P.~G.}\ \bibnamefont {Lagoudakis}},\ and\ \bibinfo {author} {\bibfnamefont {H.}~\bibnamefont {Sigurdsson}},\ }\bibfield  {title} {\bibinfo {title} {Polariton vortex chern insulator},\ }\href@noop {} {\bibfield  {journal} {\bibinfo  {journal} {Optical Materials Express}\ }\textbf {\bibinfo {volume} {13}},\ \bibinfo {pages} {2550} (\bibinfo {year} {2023})}\BibitemShut {NoStop}%
\bibitem [{\citenamefont {Tosi}\ \emph {et~al.}(2012)\citenamefont {Tosi}, \citenamefont {Christmann}, \citenamefont {Berloff}, \citenamefont {Tsotsis}, \citenamefont {Gao}, \citenamefont {Hatzopoulos}, \citenamefont {Savvidis},\ and\ \citenamefont {Baumberg}}]{tosi2012geometrically}%
  \BibitemOpen
  \bibfield  {author} {\bibinfo {author} {\bibfnamefont {G.}~\bibnamefont {Tosi}}, \bibinfo {author} {\bibfnamefont {G.}~\bibnamefont {Christmann}}, \bibinfo {author} {\bibfnamefont {N.}~\bibnamefont {Berloff}}, \bibinfo {author} {\bibfnamefont {P.}~\bibnamefont {Tsotsis}}, \bibinfo {author} {\bibfnamefont {T.}~\bibnamefont {Gao}}, \bibinfo {author} {\bibfnamefont {Z.}~\bibnamefont {Hatzopoulos}}, \bibinfo {author} {\bibfnamefont {P.}~\bibnamefont {Savvidis}},\ and\ \bibinfo {author} {\bibfnamefont {J.}~\bibnamefont {Baumberg}},\ }\bibfield  {title} {\bibinfo {title} {Geometrically locked vortex lattices in semiconductor quantum fluids},\ }\href@noop {} {\bibfield  {journal} {\bibinfo  {journal} {Nature communications}\ }\textbf {\bibinfo {volume} {3}},\ \bibinfo {pages} {1243} (\bibinfo {year} {2012})}\BibitemShut {NoStop}%
\bibitem [{\citenamefont {Gao}\ \emph {et~al.}(2018)\citenamefont {Gao}, \citenamefont {Egorov}, \citenamefont {Estrecho}, \citenamefont {Winkler}, \citenamefont {Kamp}, \citenamefont {Schneider}, \citenamefont {H{\"o}fling}, \citenamefont {Truscott},\ and\ \citenamefont {Ostrovskaya}}]{gao2018controlled}%
  \BibitemOpen
  \bibfield  {author} {\bibinfo {author} {\bibfnamefont {T.}~\bibnamefont {Gao}}, \bibinfo {author} {\bibfnamefont {O.~A.}\ \bibnamefont {Egorov}}, \bibinfo {author} {\bibfnamefont {E.}~\bibnamefont {Estrecho}}, \bibinfo {author} {\bibfnamefont {K.}~\bibnamefont {Winkler}}, \bibinfo {author} {\bibfnamefont {M.}~\bibnamefont {Kamp}}, \bibinfo {author} {\bibfnamefont {C.}~\bibnamefont {Schneider}}, \bibinfo {author} {\bibfnamefont {S.}~\bibnamefont {H{\"o}fling}}, \bibinfo {author} {\bibfnamefont {A.}~\bibnamefont {Truscott}},\ and\ \bibinfo {author} {\bibfnamefont {E.}~\bibnamefont {Ostrovskaya}},\ }\bibfield  {title} {\bibinfo {title} {Controlled ordering of topological charges in an exciton-polariton chain},\ }\href@noop {} {\bibfield  {journal} {\bibinfo  {journal} {Physical Review Letters}\ }\textbf {\bibinfo {volume} {121}},\ \bibinfo {pages} {225302} (\bibinfo {year} {2018})}\BibitemShut {NoStop}%
\bibitem [{\citenamefont {Alyatkin}\ \emph {et~al.}(2024)\citenamefont {Alyatkin}, \citenamefont {Mili{\'a}n}, \citenamefont {Kartashov}, \citenamefont {Sitnik}, \citenamefont {Gnusov}, \citenamefont {T{\"o}pfer}, \citenamefont {Sigurdsson},\ and\ \citenamefont {Lagoudakis}}]{alyatkin2024antiferromagnetic}%
  \BibitemOpen
  \bibfield  {author} {\bibinfo {author} {\bibfnamefont {S.}~\bibnamefont {Alyatkin}}, \bibinfo {author} {\bibfnamefont {C.}~\bibnamefont {Mili{\'a}n}}, \bibinfo {author} {\bibfnamefont {Y.~V.}\ \bibnamefont {Kartashov}}, \bibinfo {author} {\bibfnamefont {K.~A.}\ \bibnamefont {Sitnik}}, \bibinfo {author} {\bibfnamefont {I.}~\bibnamefont {Gnusov}}, \bibinfo {author} {\bibfnamefont {J.~D.}\ \bibnamefont {T{\"o}pfer}}, \bibinfo {author} {\bibfnamefont {H.}~\bibnamefont {Sigurdsson}},\ and\ \bibinfo {author} {\bibfnamefont {P.~G.}\ \bibnamefont {Lagoudakis}},\ }\bibfield  {title} {\bibinfo {title} {Antiferromagnetic ising model in a triangular vortex lattice of quantum fluids of light},\ }\href@noop {} {\bibfield  {journal} {\bibinfo  {journal} {Science Advances}\ }\textbf {\bibinfo {volume} {10}},\ \bibinfo {pages} {eadj1589} (\bibinfo {year} {2024})}\BibitemShut {NoStop}%
\bibitem [{\citenamefont {Cookson}\ \emph {et~al.}(2021)\citenamefont {Cookson}, \citenamefont {Kalinin}, \citenamefont {Sigurdsson}, \citenamefont {T{\"o}pfer}, \citenamefont {Alyatkin}, \citenamefont {Silva}, \citenamefont {Langbein}, \citenamefont {Berloff},\ and\ \citenamefont {Lagoudakis}}]{cookson2021geometric}%
  \BibitemOpen
  \bibfield  {author} {\bibinfo {author} {\bibfnamefont {T.}~\bibnamefont {Cookson}}, \bibinfo {author} {\bibfnamefont {K.}~\bibnamefont {Kalinin}}, \bibinfo {author} {\bibfnamefont {H.}~\bibnamefont {Sigurdsson}}, \bibinfo {author} {\bibfnamefont {J.~D.}\ \bibnamefont {T{\"o}pfer}}, \bibinfo {author} {\bibfnamefont {S.}~\bibnamefont {Alyatkin}}, \bibinfo {author} {\bibfnamefont {M.}~\bibnamefont {Silva}}, \bibinfo {author} {\bibfnamefont {W.}~\bibnamefont {Langbein}}, \bibinfo {author} {\bibfnamefont {N.~G.}\ \bibnamefont {Berloff}},\ and\ \bibinfo {author} {\bibfnamefont {P.~G.}\ \bibnamefont {Lagoudakis}},\ }\bibfield  {title} {\bibinfo {title} {Geometric frustration in polygons of polariton condensates creating vortices of varying topological charge},\ }\href@noop {} {\bibfield  {journal} {\bibinfo  {journal} {Nature communications}\ }\textbf {\bibinfo {volume} {12}},\ \bibinfo {pages} {2120} (\bibinfo {year} {2021})}\BibitemShut {NoStop}%
\bibitem [{\citenamefont {Kwon}\ \emph {et~al.}(2019)\citenamefont {Kwon}, \citenamefont {Oh}, \citenamefont {Gong}, \citenamefont {Kim}, \citenamefont {Kang}, \citenamefont {Kang}, \citenamefont {Song}, \citenamefont {Choi},\ and\ \citenamefont {Cho}}]{kwon2019direct}%
  \BibitemOpen
  \bibfield  {author} {\bibinfo {author} {\bibfnamefont {M.-S.}\ \bibnamefont {Kwon}}, \bibinfo {author} {\bibfnamefont {B.~Y.}\ \bibnamefont {Oh}}, \bibinfo {author} {\bibfnamefont {S.-H.}\ \bibnamefont {Gong}}, \bibinfo {author} {\bibfnamefont {J.-H.}\ \bibnamefont {Kim}}, \bibinfo {author} {\bibfnamefont {H.~K.}\ \bibnamefont {Kang}}, \bibinfo {author} {\bibfnamefont {S.}~\bibnamefont {Kang}}, \bibinfo {author} {\bibfnamefont {J.~D.}\ \bibnamefont {Song}}, \bibinfo {author} {\bibfnamefont {H.}~\bibnamefont {Choi}},\ and\ \bibinfo {author} {\bibfnamefont {Y.-H.}\ \bibnamefont {Cho}},\ }\bibfield  {title} {\bibinfo {title} {Direct transfer of light’s orbital angular momentum onto a nonresonantly excited polariton superfluid},\ }\href@noop {} {\bibfield  {journal} {\bibinfo  {journal} {Physical review letters}\ }\textbf {\bibinfo {volume} {122}},\ \bibinfo {pages} {045302} (\bibinfo {year} {2019})}\BibitemShut {NoStop}%
\bibitem [{\citenamefont {Coullet}\ and\ \citenamefont {Emilsson}(1992)}]{coullet1992strong}%
  \BibitemOpen
  \bibfield  {author} {\bibinfo {author} {\bibfnamefont {P.}~\bibnamefont {Coullet}}\ and\ \bibinfo {author} {\bibfnamefont {K.}~\bibnamefont {Emilsson}},\ }\bibfield  {title} {\bibinfo {title} {Strong resonances of spatially distributed oscillators: a laboratory to study patterns and defects},\ }\href@noop {} {\bibfield  {journal} {\bibinfo  {journal} {Physica D: Nonlinear Phenomena}\ }\textbf {\bibinfo {volume} {61}},\ \bibinfo {pages} {119} (\bibinfo {year} {1992})}\BibitemShut {NoStop}%
\bibitem [{\citenamefont {Coullet}\ \emph {et~al.}(1990)\citenamefont {Coullet}, \citenamefont {Lega}, \citenamefont {Houchmandzadeh},\ and\ \citenamefont {Lajzerowicz}}]{coullet1990breaking}%
  \BibitemOpen
  \bibfield  {author} {\bibinfo {author} {\bibfnamefont {P.}~\bibnamefont {Coullet}}, \bibinfo {author} {\bibfnamefont {J.}~\bibnamefont {Lega}}, \bibinfo {author} {\bibfnamefont {B.}~\bibnamefont {Houchmandzadeh}},\ and\ \bibinfo {author} {\bibfnamefont {J.}~\bibnamefont {Lajzerowicz}},\ }\bibfield  {title} {\bibinfo {title} {Breaking chirality in nonequilibrium systems},\ }\href@noop {} {\bibfield  {journal} {\bibinfo  {journal} {Physical review letters}\ }\textbf {\bibinfo {volume} {65}},\ \bibinfo {pages} {1352} (\bibinfo {year} {1990})}\BibitemShut {NoStop}%
\bibitem [{\citenamefont {Aranson}(1995)}]{aranson1995frequency}%
  \BibitemOpen
  \bibfield  {author} {\bibinfo {author} {\bibfnamefont {I.}~\bibnamefont {Aranson}},\ }\bibfield  {title} {\bibinfo {title} {Frequency selection of spiral waves in liquid crystals},\ }\href@noop {} {\bibfield  {journal} {\bibinfo  {journal} {Physical Review E}\ }\textbf {\bibinfo {volume} {51}},\ \bibinfo {pages} {R3827} (\bibinfo {year} {1995})}\BibitemShut {NoStop}%
\bibitem [{\citenamefont {Aranson}\ \emph {et~al.}(1985)\citenamefont {Aranson}, \citenamefont {Gaponov-Grekhov},\ and\ \citenamefont {Rabinovich}}]{aranson1985development}%
  \BibitemOpen
  \bibfield  {author} {\bibinfo {author} {\bibfnamefont {I.}~\bibnamefont {Aranson}}, \bibinfo {author} {\bibfnamefont {A.}~\bibnamefont {Gaponov-Grekhov}},\ and\ \bibinfo {author} {\bibfnamefont {M.}~\bibnamefont {Rabinovich}},\ }\bibfield  {title} {\bibinfo {title} {Development of chaos in ensembles of dynamical structures},\ }\href@noop {} {\bibfield  {journal} {\bibinfo  {journal} {Zh. Eksp. Teor. Fiz.}\ }\textbf {\bibinfo {volume} {89}},\ \bibinfo {pages} {92} (\bibinfo {year} {1985})}\BibitemShut {NoStop}%
\bibitem [{\citenamefont {Wouters}\ and\ \citenamefont {Carusotto}(2007)}]{wouters2007excitations}%
  \BibitemOpen
  \bibfield  {author} {\bibinfo {author} {\bibfnamefont {M.}~\bibnamefont {Wouters}}\ and\ \bibinfo {author} {\bibfnamefont {I.}~\bibnamefont {Carusotto}},\ }\bibfield  {title} {\bibinfo {title} {Excitations in a nonequilibrium bose-einstein condensate of exciton polaritons},\ }\href@noop {} {\bibfield  {journal} {\bibinfo  {journal} {Physical review letters}\ }\textbf {\bibinfo {volume} {99}},\ \bibinfo {pages} {140402} (\bibinfo {year} {2007})}\BibitemShut {NoStop}%
\bibitem [{\citenamefont {Solnyshkov}\ \emph {et~al.}(2014)\citenamefont {Solnyshkov}, \citenamefont {Ter{\c{c}}as}, \citenamefont {Dini},\ and\ \citenamefont {Malpuech}}]{solnyshkov2014hybrid}%
  \BibitemOpen
  \bibfield  {author} {\bibinfo {author} {\bibfnamefont {D.}~\bibnamefont {Solnyshkov}}, \bibinfo {author} {\bibfnamefont {H.}~\bibnamefont {Ter{\c{c}}as}}, \bibinfo {author} {\bibfnamefont {K.}~\bibnamefont {Dini}},\ and\ \bibinfo {author} {\bibfnamefont {G.}~\bibnamefont {Malpuech}},\ }\bibfield  {title} {\bibinfo {title} {Hybrid boltzmann--gross-pitaevskii theory of bose-einstein condensation and superfluidity in open driven-dissipative systems},\ }\href@noop {} {\bibfield  {journal} {\bibinfo  {journal} {Physical Review A}\ }\textbf {\bibinfo {volume} {89}},\ \bibinfo {pages} {033626} (\bibinfo {year} {2014})}\BibitemShut {NoStop}%
\bibitem [{\citenamefont {Ma}\ and\ \citenamefont {Schumacher}(2018)}]{ma2018vortex}%
  \BibitemOpen
  \bibfield  {author} {\bibinfo {author} {\bibfnamefont {X.}~\bibnamefont {Ma}}\ and\ \bibinfo {author} {\bibfnamefont {S.}~\bibnamefont {Schumacher}},\ }\bibfield  {title} {\bibinfo {title} {Vortex multistability and bessel vortices in polariton condensates},\ }\href@noop {} {\bibfield  {journal} {\bibinfo  {journal} {Physical Review Letters}\ }\textbf {\bibinfo {volume} {121}},\ \bibinfo {pages} {227404} (\bibinfo {year} {2018})}\BibitemShut {NoStop}%
\bibitem [{\citenamefont {Mu{\~n}oz~Mateo}\ \emph {et~al.}(2020)\citenamefont {Mu{\~n}oz~Mateo}, \citenamefont {Rubo},\ and\ \citenamefont {Toikka}}]{munoz2020long}%
  \BibitemOpen
  \bibfield  {author} {\bibinfo {author} {\bibfnamefont {A.}~\bibnamefont {Mu{\~n}oz~Mateo}}, \bibinfo {author} {\bibfnamefont {Y.}~\bibnamefont {Rubo}},\ and\ \bibinfo {author} {\bibfnamefont {L.}~\bibnamefont {Toikka}},\ }\bibfield  {title} {\bibinfo {title} {Long josephson junctions with exciton-polariton condensates},\ }\href@noop {} {\bibfield  {journal} {\bibinfo  {journal} {Physical Review B}\ }\textbf {\bibinfo {volume} {101}},\ \bibinfo {pages} {184509} (\bibinfo {year} {2020})}\BibitemShut {NoStop}%
\bibitem [{\citenamefont {Cherotchenko}\ \emph {et~al.}(2021)\citenamefont {Cherotchenko}, \citenamefont {Sigurdsson}, \citenamefont {Askitopoulos},\ and\ \citenamefont {Nalitov}}]{cherotchenko2021optically}%
  \BibitemOpen
  \bibfield  {author} {\bibinfo {author} {\bibfnamefont {E.}~\bibnamefont {Cherotchenko}}, \bibinfo {author} {\bibfnamefont {H.}~\bibnamefont {Sigurdsson}}, \bibinfo {author} {\bibfnamefont {A.}~\bibnamefont {Askitopoulos}},\ and\ \bibinfo {author} {\bibfnamefont {A.}~\bibnamefont {Nalitov}},\ }\bibfield  {title} {\bibinfo {title} {Optically controlled polariton condensate molecules},\ }\href@noop {} {\bibfield  {journal} {\bibinfo  {journal} {Physical Review B}\ }\textbf {\bibinfo {volume} {103}},\ \bibinfo {pages} {115309} (\bibinfo {year} {2021})}\BibitemShut {NoStop}%
\bibitem [{\citenamefont {Christmann}\ \emph {et~al.}(2012)\citenamefont {Christmann}, \citenamefont {Tosi}, \citenamefont {Berloff}, \citenamefont {Tsotsis}, \citenamefont {Eldridge}, \citenamefont {Hatzopoulos}, \citenamefont {Savvidis},\ and\ \citenamefont {Baumberg}}]{PhysRevB.85.235303}%
  \BibitemOpen
  \bibfield  {author} {\bibinfo {author} {\bibfnamefont {G.}~\bibnamefont {Christmann}}, \bibinfo {author} {\bibfnamefont {G.}~\bibnamefont {Tosi}}, \bibinfo {author} {\bibfnamefont {N.~G.}\ \bibnamefont {Berloff}}, \bibinfo {author} {\bibfnamefont {P.}~\bibnamefont {Tsotsis}}, \bibinfo {author} {\bibfnamefont {P.~S.}\ \bibnamefont {Eldridge}}, \bibinfo {author} {\bibfnamefont {Z.}~\bibnamefont {Hatzopoulos}}, \bibinfo {author} {\bibfnamefont {P.~G.}\ \bibnamefont {Savvidis}},\ and\ \bibinfo {author} {\bibfnamefont {J.~J.}\ \bibnamefont {Baumberg}},\ }\bibfield  {title} {\bibinfo {title} {Polariton ring condensates and sunflower ripples in an expanding quantum liquid},\ }\href {https://doi.org/10.1103/PhysRevB.85.235303} {\bibfield  {journal} {\bibinfo  {journal} {Phys. Rev. B}\ }\textbf {\bibinfo {volume} {85}},\ \bibinfo {pages} {235303} (\bibinfo {year} {2012})}\BibitemShut {NoStop}%
\bibitem [{\citenamefont {Sch{\"o}pf}\ and\ \citenamefont {Kramer}(1991)}]{schopf1991small}%
  \BibitemOpen
  \bibfield  {author} {\bibinfo {author} {\bibfnamefont {W.}~\bibnamefont {Sch{\"o}pf}}\ and\ \bibinfo {author} {\bibfnamefont {L.}~\bibnamefont {Kramer}},\ }\bibfield  {title} {\bibinfo {title} {Small-amplitude periodic and chaotic solutions of the complex ginzburg-landau equation for a subcritical bifurcation},\ }\href@noop {} {\bibfield  {journal} {\bibinfo  {journal} {Physical review letters}\ }\textbf {\bibinfo {volume} {66}},\ \bibinfo {pages} {2316} (\bibinfo {year} {1991})}\BibitemShut {NoStop}%
\bibitem [{\citenamefont {Kuramoto}\ and\ \citenamefont {Kuramoto}(1984)}]{kuramoto1984chemical}%
  \BibitemOpen
  \bibfield  {author} {\bibinfo {author} {\bibfnamefont {Y.}~\bibnamefont {Kuramoto}}\ and\ \bibinfo {author} {\bibfnamefont {Y.}~\bibnamefont {Kuramoto}},\ }\href@noop {} {\emph {\bibinfo {title} {Chemical turbulence}}}\ (\bibinfo  {publisher} {Springer},\ \bibinfo {year} {1984})\BibitemShut {NoStop}%
\end{thebibliography}%
 
\end{document}